\documentclass[aps,twocolumn,prb,longbibliography,showpacs,floatfix,superscriptaddress]{revtex4-2}

\usepackage{graphicx,color}
\usepackage{amsfonts}
\usepackage[figuresright]{rotating}
\usepackage{amssymb}
\usepackage{bm}
\usepackage{natbib}
\usepackage{amsmath}
\usepackage{mathtools}
\usepackage{psfrag}
\usepackage{floatrow}
\usepackage{multirow}
\usepackage{tabularx}
\usepackage{textcomp}
\usepackage{units}
\usepackage{lipsum}
\usepackage{soul}
\usepackage{comment}
\usepackage{titlesec}
\usepackage{times}
\usepackage{hyperref}
\usepackage{enumitem}
\DeclareMathAlphabet\mathbfcal{OMS}{cmsy}{b}{n}

\titlespacing*{\section}
{0pt}{1ex plus .25ex}{1ex plus 1ex}
\titlespacing*{\subsection}
{0pt}{1ex plus .25ex}{1ex plus 1ex}
\titlespacing*{\subsubsection}
{0pt}{1ex plus .25ex}{1ex plus 1ex}

\titleformat{\section}
 {\fontsize{10}{15}\bfseries }
  {\thesection}
  {1em}
  {}

\titleformat{\subsection}
  {\bfseries }
  {\thesubsection}
  {2em}
  {}

\titleformat{\subsubsection}
  {\itshape}
  {\thesubsubsection}
  {2em}
  {}

\def\beq{\begin{eqnarray}}
\def\eeq{\end{eqnarray}}

\let\baraccent=\= 
\renewcommand{\=}[1]{\stackrel{#1}{=}} 
\newcommand{\mc}[1]{\mathcal{ #1}} 

\makeatletter

\titleclass{\subsubsubsection}{straight}[\subsection]

\newcounter{subsubsubsection}[subsubsection]
\renewcommand\thesubsubsubsection{\thesubsubsection.\arabic{subsubsubsection}}

\titleformat{\subsubsubsection}
   {\normalfont\normalsize\itshape \centering}{\thesubsubsubsection}{1em}{}
\titlespacing*{\subsubsubsection}
{0pt}{1ex plus .3ex}{1ex plus .3ex}

\makeatletter
\renewcommand\paragraph{\@startsection{paragraph}{5}{\z@}%
  {3.25ex \@plus1ex \@minus.2ex}%
  {-1em}%
  {\normalfont\normalsize}}
\renewcommand\subparagraph{\@startsection{subparagraph}{6}{\parindent}%
  {3.25ex \@plus1ex \@minus .2ex}%
  {-1em}%
  {\normalfont\normalsize}}
\def\toclevel@subsubsubsection{4}
\def\toclevel@paragraph{5}
\def\toclevel@paragraph{6}
\def\l@subsubsubsection{\@dottedtocline{4}{7em}{4em}}
\def\l@paragraph{\@dottedtocline{5}{10em}{5em}}
\def\l@subparagraph{\@dottedtocline{6}{14em}{6em}}
\makeatother

\begin{document}
\title{Signatures of the quantum skyrmion Hall effect in the Bernevig-Hughes-Zhang model}
\author{Reyhan Ay}
\affiliation{Max Planck Institute for Chemical Physics of Solids, Nöthnitzer Strasse 40, 01187 Dresden, Germany}
\affiliation{Max Planck Institute for the Physics of Complex Systems, Nöthnitzer Strasse 38, 01187 Dresden, Germany}

\author{Adipta Pal}
\affiliation{Max Planck Institute for Chemical Physics of Solids, Nöthnitzer Strasse 40, 01187 Dresden, Germany}
\affiliation{Max Planck Institute for the Physics of Complex Systems, Nöthnitzer Strasse 38, 01187 Dresden, Germany}

\author{Ashley M. Cook}
\affiliation{Max Planck Institute for Chemical Physics of Solids, Nöthnitzer Strasse 40, 01187 Dresden, Germany}
\affiliation{Max Planck Institute for the Physics of Complex Systems, Nöthnitzer Strasse 38, 01187 Dresden, Germany}

\begin{abstract}
Given recent discovery of the quantum skyrmion Hall effect, we re-examine the related canonical Bernevig-Hughes-Zhang (BHZ) model for the quantum spin Hall insulator. Within the framework of the quantum skyrmion Hall effect, isospin degree(s) of freedom of the BHZ model encode additional spatial dimensions. Consistent with this framework, we observe phenomena similar to those of the four dimensional Chern insulator, revealed by weakly breaking time-reversal symmetry. Bulk-boundary correspondence of these states includes real-space boundary orbital angular momentum textures and gapless boundary modes that are robust against magnetic disorder, consistent with compactified three dimensional boundary Weyl nodes (WN\textsubscript{F}s) of the quantum skyrmion Hall effect. These theoretical findings are furthermore consistent with past experimental work reporting unexpected edge conduction in HgTe quantum wells under applied Zeeman and orbital magnetic fields. This past work is therefore potentially the first known experimental observation of signatures of the quantum skyrmion Hall effect beyond the quantum Hall effect.
\end{abstract}
\maketitle

Discovery of the quantum Hall effect (QHE)~\cite{PhysRevLett.45.494, PhysRevLett.48.1559} revealed the essential role of topology in condensed matter physics~\cite{laughlin1981,halperin1982quantized, Haldane:1983xm, niu1985, PhysRevLett.50.1395, PhysRevLett.62.82, PhysRevLett.63.199, PhysRevLett.64.1313}, providing a rich platform for experimental study. Practical applications of topological states appeared more promising, however, following discovery of topological insulators~\cite{kane2005, fu_topological_2007, moore2007}, which are topological states that can be realised without external magnetic fields or even magnetic order. Introduction of the time-reversal-invariant (TRI) quantum spin Hall insulator (QSHI) state~\cite{QSHI-HgTe-Theory} and experimental observation in HgTe QWs~\cite{koenig2007} supported theoretically by the Bernevig-Hughes-Zhang (BHZ) model, in particular, ushered in a new era in experimental study~\cite{hsieh_topological_2008, chen2009, chen2010, peng_aharonovbohm_2010, liu2014, liu_stable_2014, xu2015, tang_quantum_2017, nadjperge2014, roushan_topological_2009, xia_observation_2009, huang_weyl_2015}, motivating intense efforts to harness these states for applications~\cite{KITAEV20032, KITAEV20062, read2000, nayak2008}.

Considerable progress has been made to map out the landscape of topologically non-trivial phases of matter since these pioneering efforts~\cite{Ryu_2010, schnyder2008, kitaev_periodic_2009}, recently leading to discovery of a generalisation from the framework of the QHE to that of the quantum skyrmion Hall effect (QSkHE)~\cite{qskhe, patil2024}. If $\delta$+$1$-dimensional ($\delta$+$1$ D, where $+1$ is the time dimension) topologically non-trivial many-body states are compactified (made almost point-like by dimensional reduction schemes~\cite{patil2024, bernevig6Dfieldtheory}), they remain intrinsically $\delta$+$1$ D many-body states, but encoded in isospin degrees of freedom (DOFs). These compactified states play the role, in the QSkHE, that charged (quasi)particles play in the QHE, generalising notions of incompressible states. 

Many phenomena of the QSkHE beyond the QHE framework have recently been identified theoretically~\cite{cook_multiplicative_2022, cook2023, cookFST2023,liu2020, qskhe, pal_multsemimetal, pal2024_fstwsm, winterOEPT, calderon_skyrm, calderon2023_fst, calderongapless2023, ay2023, pacholski2024, liu2023skyrmsemimetals}, motivating a minimal effective field theory of the QSkHE as a generalisation of the 2+1 D SU(2) gauge theory by Patil~\emph{et al.}~\cite{patil2024}. The BHZ model, being related to the 2+1 D SU(2) gauge theory~\cite{qi2008TRIFT}, is therefore also expected to host signatures of the QSkHE, which we characterise in this work. Especially significant is a previously-unidentified bulk-boundary correspondence of a compactified 4D Chern insulator (CI) state of the QSkHE framework, consisting of compactified 3D Weyl nodes that are exceptionally robust against time-reversal symmetry-breaking perturbations. 

To later discuss the BHZ model, we first review a minimal action for the QSkHE in terms of a mixture of two Cartesian spatial coordinates and two extra fuzzy dimensions from separate work~\cite{patil2024},
\begin{align}
    S_{CS}=&C_2\int d^3x k\;Tr\left[Tr(G)\;CS_3\wedge \Hat{F}\right],\\
   CS_3=&\left[AdA+\frac{2}{3}AAA\right],\\
   \Hat{F}= & \left[ X_a, A_b \right]   - \left[ X_b, A_a \right] + \left[A_a, A_b \right] - C^c_{ab} A_c,
\end{align}
where $C_2$ is the second Chern number, $CS_3$ is the standard 2+1 D non-Abelian Chern-Simons Lagrange density, and $\hat{F}$ is the field strength defined over a fuzzy coset space $\left(SO(5)/SU(2) \right)_F \cong S^2_F$, where $S^2_F$ is the fuzzy two-sphere~\cite{qskhe, aschieri2007, Aschieri:2003vy, Aschieri:2004vh, patil2024}, expressed in terms of $N \times N$ matrix Lie algebra generators $\{X_a\}$, gauge field components $\{A_a\}$, and structure constant $C^c_{ab}$. $Tr(G)$ is the trace over the gauge group $G$, and $kTr$ is the normalised trace over fuzzy coset space coordinates. From this perspective, the minimal phenomenology of the QSkHE relevant to the BHZ model is that of the 4+1 D QHE~\cite{zhanghu2001, bernevig6Dfieldtheory} and 4+1 D CI~\cite{qi2008TRIFT} in higher symmetry cases~\cite{patil2024}, but with replacement of Landau levels (LLs) by \textit{severely-fuzzified Landau levels}, or LL\textsubscript{F}s~\cite{patil2024}. In the simplest cases, the LL\textsubscript{F} is a LL defined in terms of position coordinates proportional to $N \times N$ matrix Lie algebra generators, where $N$ is small ($N=4$ in this work), and the LL\textsubscript{F} is almost point-like. The key generalisation of the QSkHE, however, is that the many-body state defined by filling some number of these orbitals of the LL\textsubscript{F} \textit{is still intrinsically 2+1 D}~\cite{qskhe, patil2024}. Propagating LL\textsubscript{F}s at the boundary are also understood as severely-fuzzified 3D Weyl nodes, or WN\textsubscript{F}s~\cite{qi2008TRIFT}.

\textit{Hamiltonian}---We consider the Bernevig-Hughes-Zhang (BHZ) model~\cite{qi2008TRIFT, QSHI-HgTe-Theory}, with Hamiltonian 
\begin{align}
\mathcal{H} = \sum_{\boldsymbol{k}} \psi^{\dagger}_{\boldsymbol{k}}H(\boldsymbol{k}) \psi^{}_{\boldsymbol{k}},
\end{align}
in terms of Bloch Hamiltonian $H(\boldsymbol{k})$, with momentum $\boldsymbol{k} = \left(k_x, k_y \right)$, and basis vector $\psi^{\dagger}_{\boldsymbol{k}} = \left(c^{\dagger}_{\boldsymbol{k}, \uparrow, \alpha}, c^{\dagger}_{\boldsymbol{k},\uparrow, \beta}, c^{\dagger}_{\boldsymbol{k}, \downarrow, \alpha}, c^{\dagger}_{\boldsymbol{k}, \downarrow, \beta}\right)$, where $c^{\dagger}_{\boldsymbol{k}, \sigma, \ell}$ creates a fermion with momentum $\boldsymbol{k}$, spin $\sigma \in \left\{ \uparrow, \downarrow \right\}$, and orbital angular momentum (OAM) $\ell \in \left\{\alpha, \beta \right\}$. In this basis, the Bloch Hamiltonian $H(\boldsymbol{k})$ takes the following form in terms of Kronecker products of Pauli matrices, $\tau^i \sigma^j$
\begin{align}
    H(\boldsymbol{k}) &= d_z(\boldsymbol{k}) \tau^0 \sigma^z + d_x(\boldsymbol{k}) \tau^z \sigma^x + d_y(\boldsymbol{k}) \tau^0 \sigma^y \\ \nonumber
    &+ c_A \tau^x \sigma^y + c_R (f_x\left(\boldsymbol{k}\right)\tau^x \sigma^0 + f_y\left(\boldsymbol{k}\right)\tau^y \sigma^0)+ \boldsymbol{h}\cdot \boldsymbol{\tau}\sigma^0.
\end{align}

Here, the $\boldsymbol{d}$-vector $\boldsymbol{d}(\boldsymbol{k}) = \left(d_x(\boldsymbol{k}), d_y(\boldsymbol{k}), d_z(\boldsymbol{k}) \right)$ is taken to have components as $d_x(\boldsymbol{k})= \sin(k_x)$, $d_y(\boldsymbol{k})= \sin(k_y)$, and $d_z(\boldsymbol{k})= u + \cos(k_x) + \cos(k_y)$, and $u$ is as an effective mass. The block-diagonal of $H(\boldsymbol{k})$ consists of two Bloch Hamiltonians related by time-reversal symmetry (TRS), each an instance of the Qi-Wu-Zhang (QWZ) model for a CI~
\cite{qi2006_QWZmodel}. $c_A$ denotes atomic spin-orbit coupling (SOC) strength and $c_R$ denotes Rashba SOC strength.  $\boldsymbol{f}$ is a vector $\boldsymbol{f}(\boldsymbol{k}) = \left(f_x(\boldsymbol{k}), f_y(\boldsymbol{k}), f_z(\boldsymbol{k}) \right)$ with components $f_x(\boldsymbol{k})= \sin(k_y)$, $f_y(\boldsymbol{k})= -\sin(k_x)$ and $f_z(\boldsymbol{k})= 0$. $\boldsymbol{h}=\left(h_x, h_y, h_z \right)$ encodes an effective Zeeman field, where $\boldsymbol{\tau} = \left(\tau^x, \tau^y, \tau^z \right)$ is a vector of Pauli matrices.

For finite atomic SOC and finite Zeeman field strength oriented in the out-of-plane $\hat{z}$-direction, with negligible Rashba SOC, the topological classification of the QSHI is $\nu \in \mathbb{Z}_2$~\cite{kane2005}. The topological phase diagram in this case is known and closely tied to that of the underlying QWZ model. In the $u-c_A$ plane, $\nu$ is non-trivial in two circular regions centered at $u = \pm 1$ and $c_A=0$, with unit radius~\cite{calderon2023_fst}.

\textit{Orbital angular momentum textures in momentum-space and real-space for finite Zeeman field}---To characterise signatures of the quantum skyrmion Hall effect (QSkHE) in the BHZ model, we introduce the skyrmion number~\cite{cook2023} as
\begin{align}
    \mathcal{Q} = {1 \over 4\pi}\int_{BZ} d\boldsymbol{k} \langle \boldsymbol{S} \left(\boldsymbol{k} \right)  \rangle \cdot \left(\partial_{k_x}\langle \boldsymbol{S} \left(\boldsymbol{k} 
\rangle \right) \times \partial_{k_y}\langle \boldsymbol{S} \left(\boldsymbol{k} \right) \rangle \right),
\end{align}
where $ \langle \boldsymbol{S} \left(\boldsymbol{k} \right)  \rangle = \left(\langle S_x \left(\boldsymbol{k} \right)  \rangle, \langle S_y \left(\boldsymbol{k} \right)  \rangle, \langle S_z \left(\boldsymbol{k} \right)  \rangle \right) $ is the ground-state OAM expectation value computed as $\langle S_i \left(\boldsymbol{k} \right)  \rangle = \sum_n^{occ} \langle n, \boldsymbol{k}| S_i |n, \boldsymbol{k}\rangle$, where $|n, \boldsymbol{k}\rangle$ is the Bloch eigenstate for the $n$\textsuperscript{th} lowest-energy band at momentum $\boldsymbol{k}$, and $n$ runs over occupied states. Here, we assume half-filling of the bands. Unless stated otherwise, the matrix representations of the OAM operators are taken to be $S_x = \tau^z \sigma^x$, $S_y = \tau^0 \sigma^y$, and $S_z = \tau^0 \sigma^z$.

\begin{figure}
   \centering
   \includegraphics[width=\linewidth]{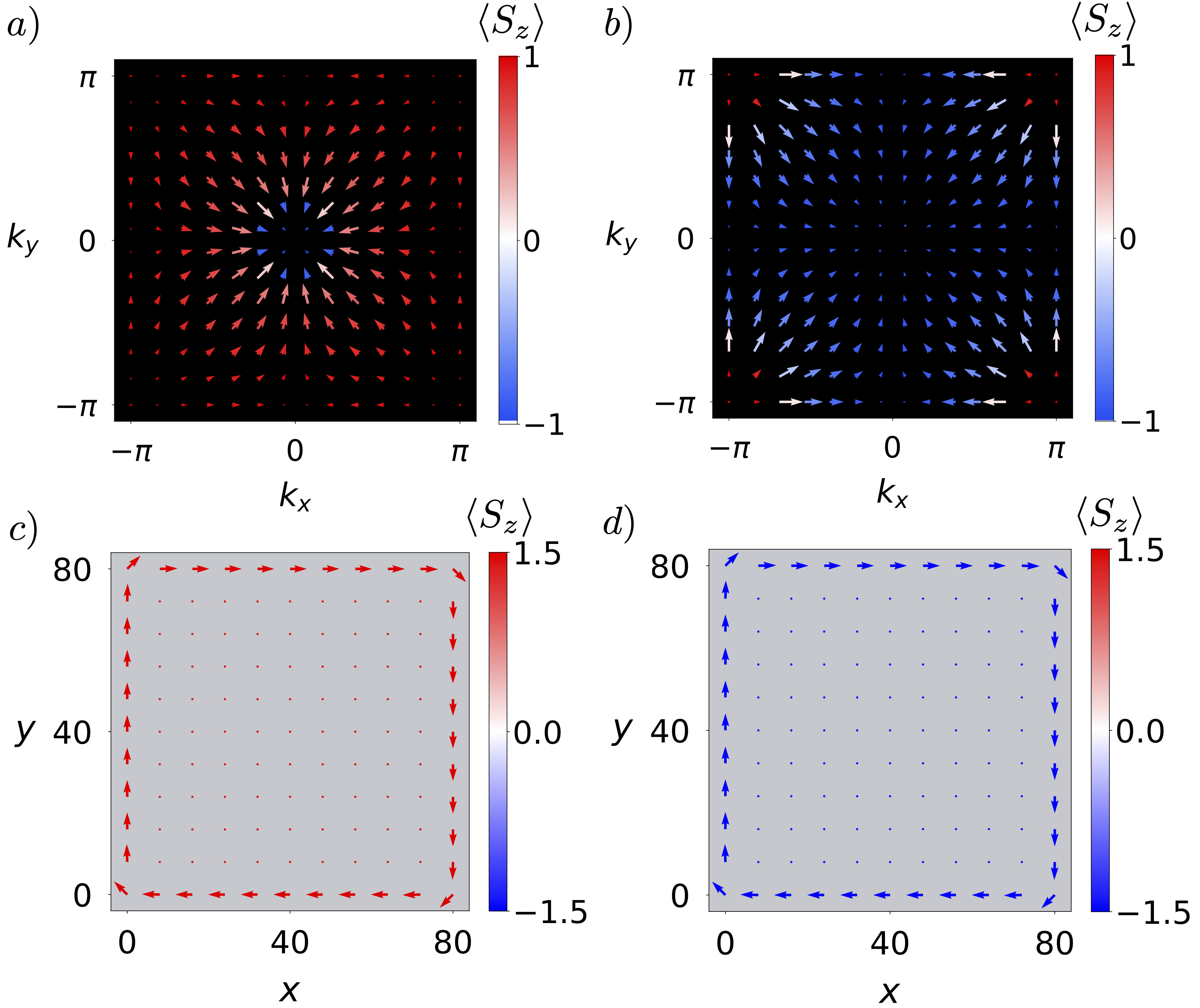}
    \caption{Normalized momentum-space ground-state OAM expectation value textures $\langle S(\boldsymbol{k})\rangle$ for a) $u=-1.56$, $c_A=-0.77$, $\boldsymbol{h}=0.14\hat{z}$, with $\mathcal{Q}=-1$ and b) $u=1.56$, $c_A=-0.77$, $\boldsymbol{h}=0.14\hat{z}$, with $\mathcal{Q}=+1$. Unnormalized real-space ground state OAM expectation value textures $\langle S(\boldsymbol{x,y})\rangle$ with OBCs in each of $\hat{x}$ and $\hat{y}$ directions for c) same parameters as a), and d) same parameters as in b).}
    \label{fig1}
\end{figure}

Without applied Zeeman field, $\mathcal{Q}$ is non-trivial in each region where $\nu$ is non-trivial, taking value $\pm1$ in the region centered at $u=\pm1$, respectively. Under application of a weak Zeeman field in the $\hat{z}$-direction, these values of $\mc{Q}$ remain unchanged. Examples of momentum-space skyrmionic textures associated with non-trivial $\mathcal{Q}$ are shown in Fig.~\ref{fig1} a) and b), respectively. For open boundary conditions (OBCs) in each of the $\hat{x}$- and $\hat{y}$-directions, we also study OAM textures in real-space.  For finite Zeeman field, non-trivial textures are observed and shown in Fig.~\ref{fig1} c) and d), respectively. Notably, change in sign of $\mathcal{Q}$ does not change the sign of in-plane OAM components, but does change the sign of the out-of-plane OAM component. This is a previously-observed signature of topological skyrmion phases of matter~\cite{winterOEPT}.

\begin{figure}
   \centering
   \includegraphics[width=.95\linewidth]{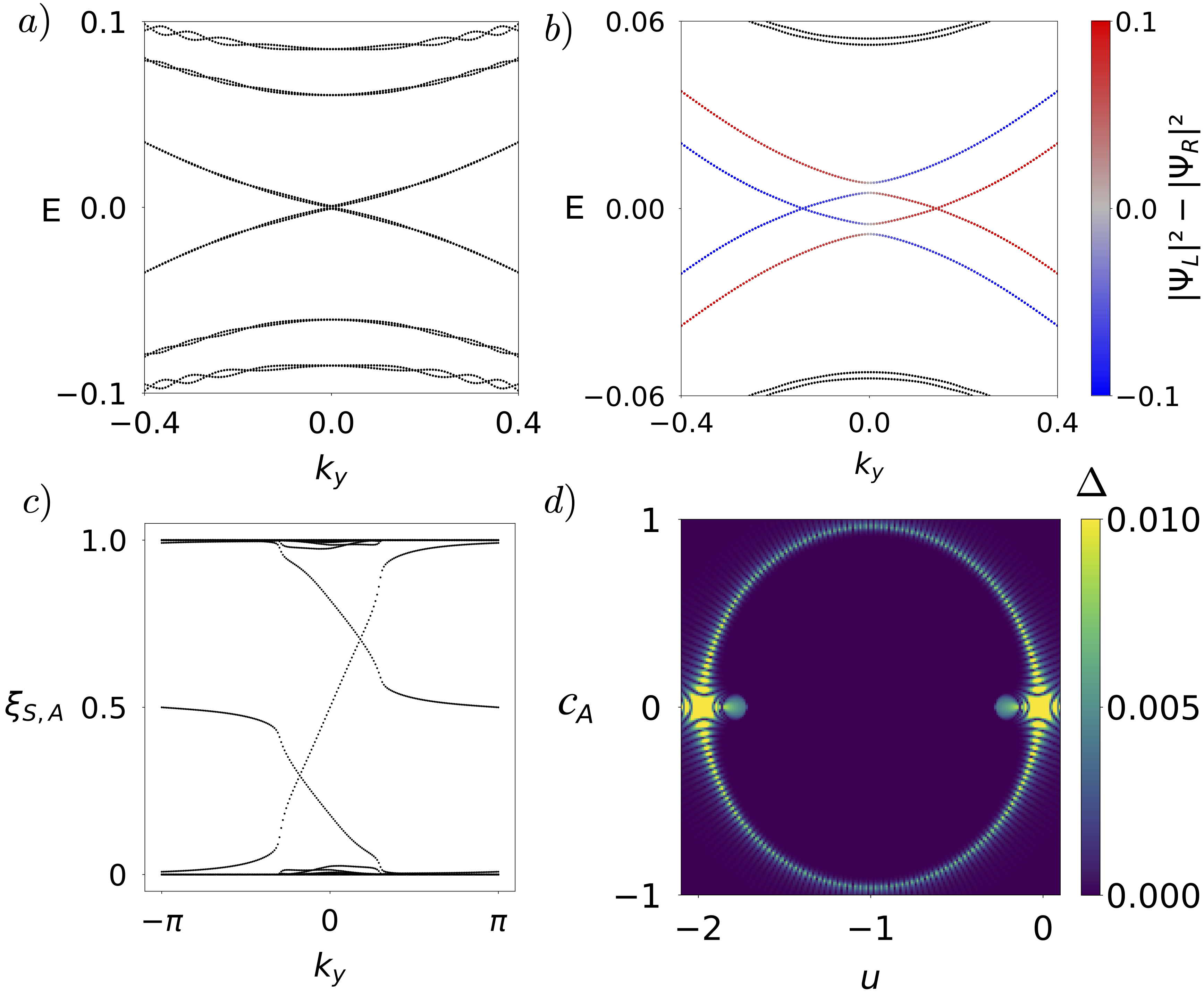}
    \caption{a) Slab spectrum for OBCs in the $\hat{x}$-direction with parameter set $u=-1.56$, $c_A=-0.77$, and $\boldsymbol{h}=0$. b) slab spectrum as in a) except $\boldsymbol{h}=0.14\hat{z}$. Color bar characterises localisation of in-gap states. c) OEES $\xi_{S,A}$ for PBCs. d) Hybridisation gap $\Delta$ between the two in-gap states with the smallest positive eigenvalues at $k_y=0$ as shown in b), plotted vs. $u$ and $c_A$.  System size in the $\hat{x}$-direction is $N=100$.}
    \label{fig2}
\end{figure}

\textit{Bulk-boundary correspondence under weak Zeeman field}---We now characterise bulk-boundary correspondence of the BHZ model consistent with the QSkHE. Bulk-boundary correspondence under TRS  is shown in Fig.~\ref{fig2} a) for OBCs in the $\hat{x}$-direction. Helical edge states are observed, which appear to be those of the QSHI. They evolve to unexpected edge states at finite Zeeman field strength, however, as shown in Fig.~\ref{fig2} b). The slab spectrum is gapless as expected at two finite $k_y$ values, but an unexpected finite hybridisation gap $\Delta$, is observed at $k_y=0$, such that individual bands in the gap \textit{delocalise} at $k_y=0$. Within the QSkHE, this hybridisation gap $\Delta$ is a signature of an intrinsically $4$+$1$ D topological phase, due to two Cartesian spatial coordinates, and two spatial dimensions encoded in the OAM operator matrix representations and Lie algebra~\cite{patil2024}. $\mathcal{Q}$ characterises this higher-dimensional topology, being similar to a second Chern number~\cite{patil2024}, and the gapless points in the slab spectrum correspond to WN\textsubscript{F}s. From this viewpoint, the finite hybridisation gap is a consequence of the Nielsen-Ninomiya theorem applicable to Weyl nodes regularised on a lattice~\cite{Nielsen:1981hk}. 

To further characterise bulk-boundary correspondence, we employ the observable-enriched (OE) partial trace (OEPT) $\Tilde{\mathrm{Tr}}$~\cite{winterOEPT}. In this method, we perform an operation on the full density matrix of the BHZ model, $\rho$, to generate a particular auxiliary density matrix, $\rho_s$. $\rho_s$ is constructed from the OAM expectation value $\langle S \rangle$ by satisfying the relation $\text{Tr}[\rho S ] = \text{Tr}[\rho_s \tilde{S}]$, where $\text{Tr}$ is the trace operation and $\tilde{S}$ is the minimal OAM operator matrix representation in the absence of the spin half DOF. This defines a generalised partial trace,  $\rho_s = \Tilde{\mathrm{Tr}}[\rho]$, which preserves the expectation value of the (pseudo)spin observable of interest, here OAM. 

We compute the OE entanglement spectrum (OEES) introduced in Winter~\emph{et al.}~\cite{winterOEPT}, shown in Fig.~\ref{fig2} c) for a system with periodic boundary conditions (PBCs) in each direction, tracing out the spin half DOF and half of the system in real-space. We observe chiral edge modes in correspondence with $\mathcal{Q}$, also indicating a topological skyrmion phase of matter~\cite{winterOEPT}. Finally, we show a color map plot of the hybridisation gap $\Delta$ in Fig.~\ref{fig2} d). The regime of finite hybridisation gap $\Delta$ nucleates in the vicinity of the $\nu \neq0$ phase boundaries. This hybridisation gap is enhanced by increasing applied Zeeman field strength and also specifically the component in the $\hat{y}$-direction. The fine-tuned gapless lines between regions with finite $\Delta$ increase in number with increasing system size, while maximum $\Delta$ decreases approximately linearly with increasing system size.  

\begin{figure}
   \centering
   \includegraphics[width=\linewidth]{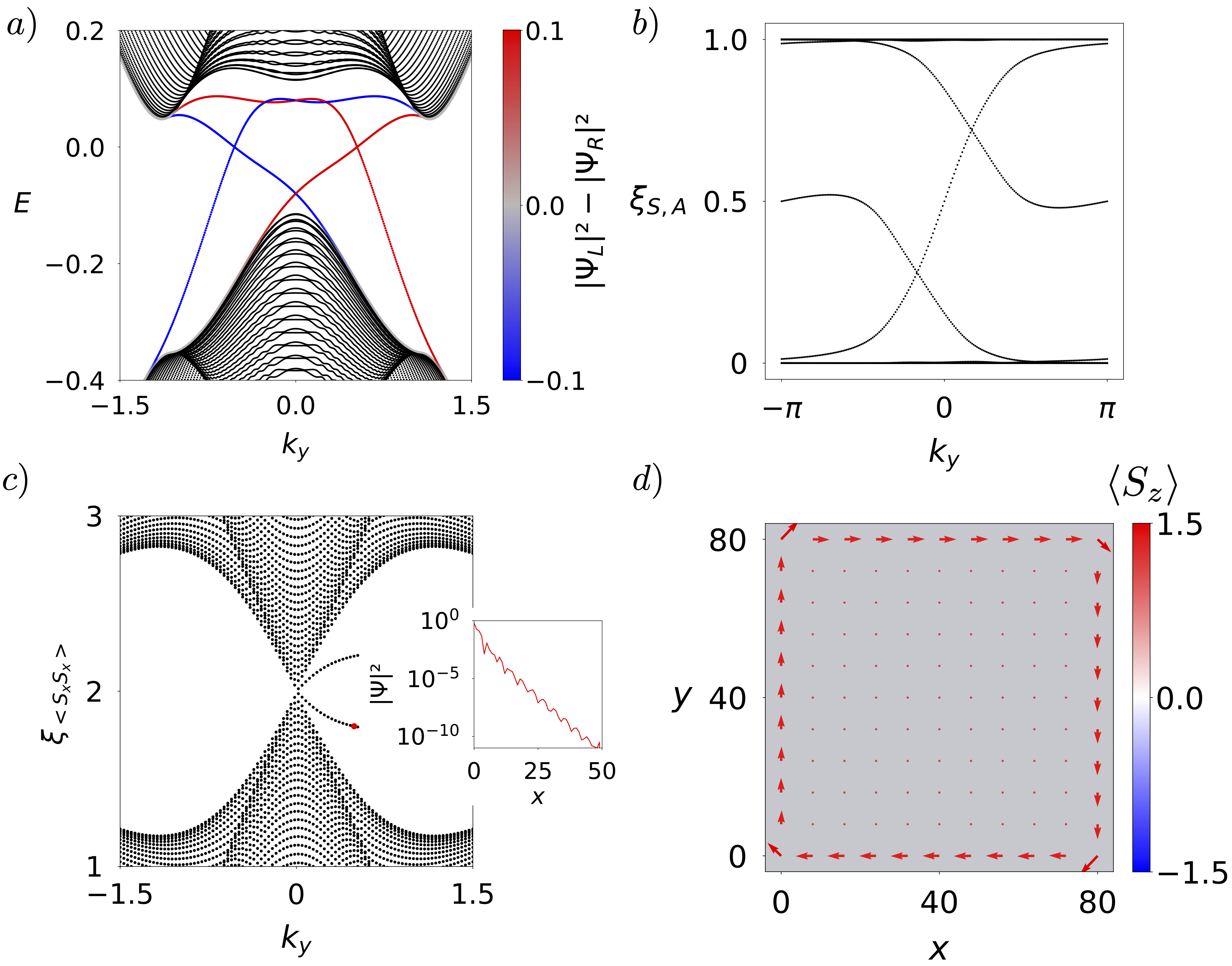}
    \caption{For parameter set $u=-1.30$, $c_A=0.50$, $c_R=0.25$, $\boldsymbol{h}=0.50\hat{z}$, a) slab spectrum for OBCs in the $\hat{x}$-direction vs. $k_y$, b) OEES $\xi_{S,A}$ for PBCs and same parameters as in a), c) spin correlator entanglement spectrum $\xi_{\langle S_x S_x \rangle}$, and d) unnormalized real-space ground state OAM expectation value texture $\langle S (x,y) \rangle$.}
    \label{fig3}
\end{figure}

\textit{Topological skyrmion phases of the BHZ model in the presence of Rashba spin-orbit coupling}---We now characterise another example of a topological skyrmion phase in the BHZ model with finite Rashba SOC. A representative slab spectrum for OBCs in the $\hat{x}$-direction is shown in Fig.~\ref{fig3} a). Importantly, despite TRS-breaking and crossing of counter-propagating bands on each edge, the spectrum is robustly gapless, even upon rotating the Zeeman field to include finite in-plane components. $\mathcal{Q}$ is also non-trivial ($\mathcal{Q}$ = -1) in this case, and the OEES exhibits a single chiral edge mode localised on each entanglement cut for the same geometry as in the case of negligible Rashba SOC in Fig.~\ref{fig3} c), in correspondence with $\mathcal{Q}$. Here, we have taken the matrix representations of the OAM operators to be $S_x = \tau^z \sigma^x + \tau^x \sigma^z$, $S_y = \tau^0 \sigma^y + \tau^y \sigma^z$, and $S_z = \tau^0 \sigma^z$, to encode inter-orbital transitions between different spin sectors, similarly to enlarged spin representations used to characterise topological skyrmion phases of matter in lower-symmetry models~\cite{cook2023, qskhe, ay2023}. 

To further characterise bulk-boundary correspondence in this case, we also compute entanglement spectra of two-point, equal-time spin correlators~\cite{qskhe}. Dividing the system in a cylinder geometry (OBCs in the $\hat{x}$-direction, PBCs in the $\hat{y}$-direction) into two halves, $A$ and $B$. We define the one-point correlator $\langle c^{\dagger}_i c^{}_j \rangle = \mathbb{P}_{\mathrm{occ}, ij}$, the $\alpha$\textsuperscript{th} OAM ($\alpha = x$, $y$, or $z$) operators for layer $i$ and $j$ in the $\hat{x}$-direction as $S^{\alpha}_i$ and $S^{\alpha}_j$, respectively, and the projector onto the $A$ half of the cylinder, $\mathbb{P}_{A}$.  Rather than computing an entanglement spectrum effectively as the spectrum of $\mathbb{P}_{A} \mathbb{P}_{\mathrm{occ}, ij} \mathbb{P}_{A}$ as in Peschel's method~\cite{IngoPeschel_2003}, we compute the spectrum of  $\mathbb{P}_{A} S^{\alpha}_i \mathbb{P}_{\mathrm{occ}, ij}  S^{\alpha}_j \mathbb{P}_{A}$.

Within the correlator entanglement spectrum, shown in Fig.~\ref{fig3} c), we observe edge states, which are exponentially-localised on the physical edge and appear within the gap in the bulk spectrum at a $k_y$ value corresponding to crossing of edge states in the energy spectrum. Such edge states are similar to those previously observed for topological skyrmion phases in three-band models~\cite{qskhe}. Finally, we also compute a real-space OAM texture for OBCs in each of the $\hat{x}$- and $\hat{y}$-directions, observing a chiral OAM texture localised at the boundary as in the cases for negligible Rashba SOC.

\textit{Unexpected edge conduction from compactified 3D Weyl nodes WN\textsubscript{F}s}---In this section, we first briefly review past experimental work on HgTe QWs~\cite{ma_unexpected_2015}. In this past work, unexpected conduction at two edges is reported for HgTe QWs realising the QSHI phase, in the presence of a finite Zeeman field and orbital magnetic field. Under these circumstances, theory of the QSHI predicts gradual loss of local density of states (LDOS) at the edge due to magnetic backscattering. Edge states are lost at a critical orbital field 
strength $B_c$ due to loss of the bulk band inversion required for the QSHI state. However, edge conduction in the experiment persisted without gradual loss of LDOS at the edge, well beyond the critical orbital magnetic field strength. This effect was not explained in this past work. 

\begin{figure}
    \centering
    \includegraphics[width=\linewidth]{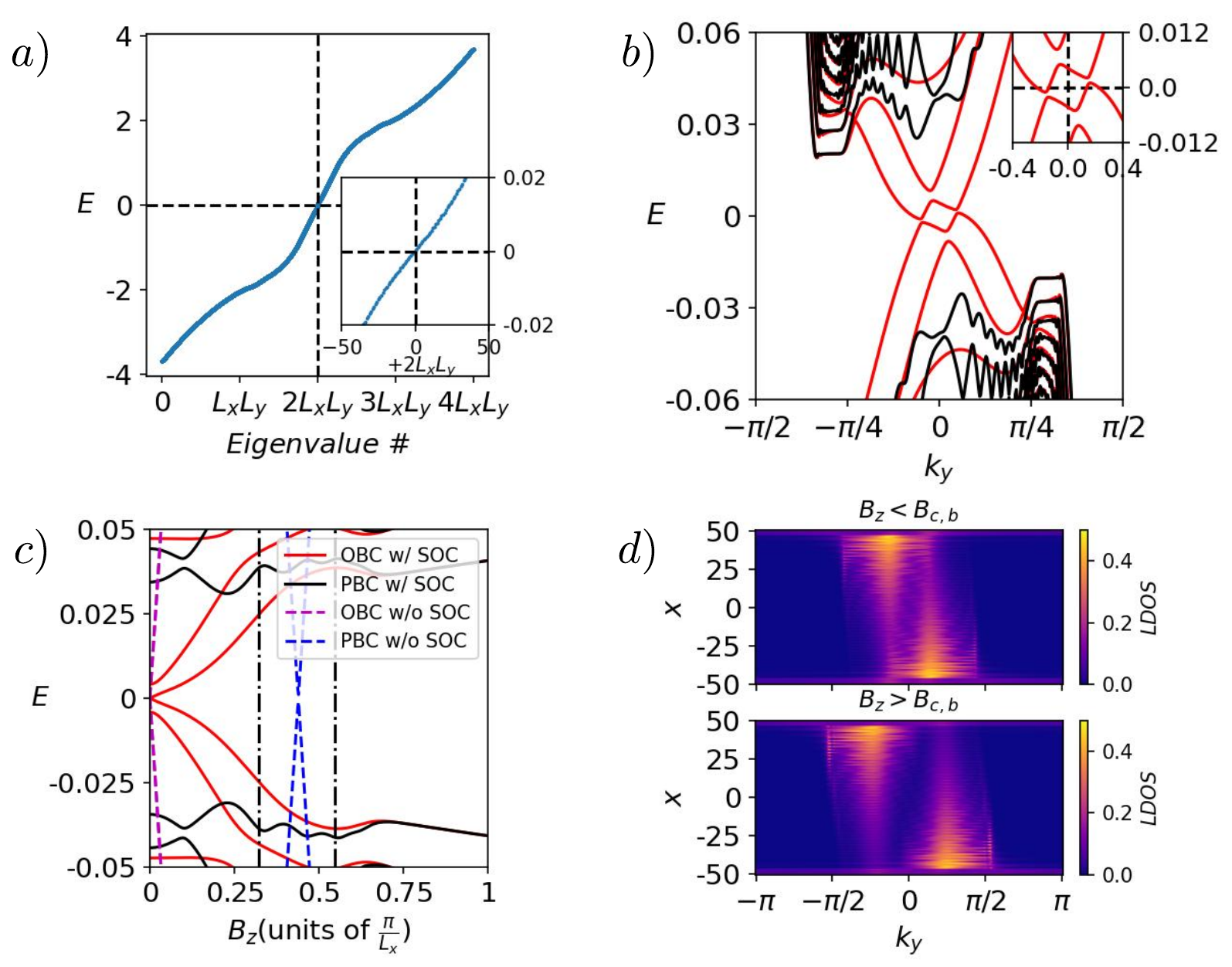}
    \caption{For parameters $u=-1.56$, and $c_A=-0.77$, a) energy eigenvalues vs. eigenvalue index for OBCs in each direction with $L_x=L_y=101$, and $h_z=0.12B_z$ with $B_z=0.003$ in presence of disorder in $\boldsymbol{h}_\parallel=(h_x,h_y)$ in the range $(-0.05,0.05)$ averaged over 50 disorder realisations with inset showing spectrum close to zero energy, b) slab spectra vs. $k_y$ with both PBCs (black) and OBCs (red) for $L_x=101$ and tilted Zeeman field with $\boldsymbol{h}=0.03\hat{x}+0.04\hat{y}+0.12B_z\hat{z}$ and $B_z=0.003$ with inset showing dispersion close to zero energy, c) energy vs. orbital out-of-plane magnetic field, $B_z$, at $k_y=0$ with PBCs (black) and OBCs in the $\hat{x}$-direction(red) and same for $c_A=0$ with PBCs (blue) and OBCs(magenta), d) LDOS for energy $E=0$ vs. $k_y$ and $x$ for $B_z=0.010$ (top) and $B_z=0.017$ (bottom) denoted by dash-dotted black vertical lines in c).}
    \label{fig4}
\end{figure}

Given discovery of the QSkHE and these past unexplained experiments, we further investigate effects of interplay between Zeeman and orbital magnetic fields on the  WN\textsubscript{F}s. We model the Zeeman field with the term $\mathcal{H}_Z= \boldsymbol{h}\cdot\boldsymbol{\tau}\sigma^0 = g\mathbf{B}\cdot\boldsymbol{\tau}\sigma^0$ as in Ma~\emph{et al.}~\cite{ma_unexpected_2015}, where $\mathbf{B}$ is the orbital field and $g$ is the g-factor. The orbital field is introduced via the gauge fixing condition for the Peierls substitution $k_y\rightarrow k_y^\prime = k_y+eB_zx$. Modeling magnetic disorder in terms of random in-plane Zeeman field components, we find the spectrum for OBCs in each of the $\hat{x}$- and $\hat{y}$-directions remains gapless for finite disorder strength, as shown in Fig.~\ref{fig4} a), consistent with interpretation of zero-energy gapless points in Fig.~\ref{fig2} b) as WN\textsubscript{F}s, given robustness of 3D Weyl nodes~\cite{wan2011}. Computing the slab spectrum for OBCs in the $\hat{x}$-direction without disorder but with weak rotation of the applied Zeeman field away from the $\hat{z}$-axis, we find the slab spectrum develops a finite minimum direct gap, but also a negative indirect gap for these OBCs, as shown in Fig.~\ref{fig4} b). The negative indirect gap is very directly associated with an anomalous effect of finite $h_x$ on the edge states: $h_x$ does not immediately gap out the slab spectrum in the regime of finite $\Delta$ as expected for helical edge states of the QSHI. Interplay between Zeeman field and orbital field very generically then yields a negative indirect gap at the edge for these OBCs, ensuring sizeable LDOS for any value of the Fermi level, see Supplementary Materials (SM) for further details~\cite{SuppMat}. 

We distinguish between orbital field strengths $B$ less than and greater than a critical value $B_c$, which corresponds to a bulk gap-closing marked by dashed blue lines, shown in Fig.~\ref{fig4} c). LDOS for $B<B_c$ is shown in Fig.~\ref{fig4} d) (top) and LDOS for $B>B_c$ is shown in Fig.~\ref{fig4} d) (bottom), indicating edge conduction persists to field strengths beyond $B_c$. We note that topological skyrmion phases of the QSkHE are known to be robust up to the magnitude of $|\langle \boldsymbol{S}(\boldsymbol{k}) \rangle |$ going to zero for some $\boldsymbol{k}$, rather than the minimum direct bulk energy gap, in general, consistent with edge conduction beyond $B_c$, which cannot be attributed to the QSHI. We note that such edge conduction associated with  WN\textsubscript{F}s can also occur more widely in phase space beyond the regions of finite hybridisation gap shown in Fig.~\ref{fig2} d) with appropriate state preparation (see SM)~\cite{SuppMat}. 

\textit{Discussion \& Conclusion}---We characterise the Bernevig-Hughes-Zhang (BHZ) model relevant to HgTe quantum wells from the perspective of the recently-discovered quantum skyrmion Hall effect (QSkHE) framework~\cite{qskhe, patil2024}. We identify signatures of topological skyrmion phases of matter~\cite{cook2023}, which are compactified 4D Chern insulators within  the QSkHE, exhibiting compactified 3D Weyl nodes (WN\textsubscript{F}s) at the boundary. Response of  WN\textsubscript{F}s to application of weak Zeeman and orbital magnetic fields is consistent with past experimental observation of unexpected edge conduction in HgTe quantum wells~\cite{ma_unexpected_2015}. We also note a later experimental work on the circular photogalvanic effect (CPGE) in HgTe QWs~\cite{dantscher2017} suggestive of boundary WN\textsubscript{F}s, given known CPGE signatures of 3D Weyl nodes~\cite{de_juan_quantized_2017}. This experimental work by Ma~\emph{et al.}~\cite{ma_unexpected_2015} therefore potentially constitutes the first known experimental observation of signatures of the QSkHE beyond the simpler framework of the quantum Hall effect.

\textit{Acknowledgements}---We would like to thank C. Xu for helpful discussions.

\bibliography{ref1.bib}

\begin{thebibliography}{63}%
\makeatletter
\providecommand \@ifxundefined [1]{%
 \@ifx{#1\undefined}
}%
\providecommand \@ifnum [1]{%
 \ifnum #1\expandafter \@firstoftwo
 \else \expandafter \@secondoftwo
 \fi
}%
\providecommand \@ifx [1]{%
 \ifx #1\expandafter \@firstoftwo
 \else \expandafter \@secondoftwo
 \fi
}%
\providecommand \natexlab [1]{#1}%
\providecommand \enquote  [1]{``#1''}%
\providecommand \bibnamefont  [1]{#1}%
\providecommand \bibfnamefont [1]{#1}%
\providecommand \citenamefont [1]{#1}%
\providecommand \href@noop [0]{\@secondoftwo}%
\providecommand \href [0]{\begingroup \@sanitize@url \@href}%
\providecommand \@href[1]{\@@startlink{#1}\@@href}%
\providecommand \@@href[1]{\endgroup#1\@@endlink}%
\providecommand \@sanitize@url [0]{\catcode `\\12\catcode `\$12\catcode
  `\&12\catcode `\#12\catcode `\^12\catcode `\_12\catcode `\%12\relax}%
\providecommand \@@startlink[1]{}%
\providecommand \@@endlink[0]{}%
\providecommand \url  [0]{\begingroup\@sanitize@url \@url }%
\providecommand \@url [1]{\endgroup\@href {#1}{\urlprefix }}%
\providecommand \urlprefix  [0]{URL }%
\providecommand \Eprint [0]{\href }%
\providecommand \doibase [0]{https://doi.org/}%
\providecommand \selectlanguage [0]{\@gobble}%
\providecommand \bibinfo  [0]{\@secondoftwo}%
\providecommand \bibfield  [0]{\@secondoftwo}%
\providecommand \translation [1]{[#1]}%
\providecommand \BibitemOpen [0]{}%
\providecommand \bibitemStop [0]{}%
\providecommand \bibitemNoStop [0]{.\EOS\space}%
\providecommand \EOS [0]{\spacefactor3000\relax}%
\providecommand \BibitemShut  [1]{\csname bibitem#1\endcsname}%
\let\auto@bib@innerbib\@empty
\bibitem [{\citenamefont {Klitzing}\ \emph {et~al.}(1980)\citenamefont
  {Klitzing}, \citenamefont {Dorda},\ and\ \citenamefont
  {Pepper}}]{PhysRevLett.45.494}%
  \BibitemOpen
  \bibfield  {author} {\bibinfo {author} {\bibfnamefont {K.~v.}\ \bibnamefont
  {Klitzing}}, \bibinfo {author} {\bibfnamefont {G.}~\bibnamefont {Dorda}},\
  and\ \bibinfo {author} {\bibfnamefont {M.}~\bibnamefont {Pepper}},\
  }\bibfield  {title} {\bibinfo {title} {New method for high-accuracy
  determination of the fine-structure constant based on quantized hall
  resistance},\ }\href {https://doi.org/10.1103/PhysRevLett.45.494} {\bibfield
  {journal} {\bibinfo  {journal} {Phys. Rev. Lett.}\ }\textbf {\bibinfo
  {volume} {45}},\ \bibinfo {pages} {494} (\bibinfo {year} {1980})}\BibitemShut
  {NoStop}%
\bibitem [{\citenamefont {Tsui}\ \emph {et~al.}(1982)\citenamefont {Tsui},
  \citenamefont {Stormer},\ and\ \citenamefont
  {Gossard}}]{PhysRevLett.48.1559}%
  \BibitemOpen
  \bibfield  {author} {\bibinfo {author} {\bibfnamefont {D.~C.}\ \bibnamefont
  {Tsui}}, \bibinfo {author} {\bibfnamefont {H.~L.}\ \bibnamefont {Stormer}},\
  and\ \bibinfo {author} {\bibfnamefont {A.~C.}\ \bibnamefont {Gossard}},\
  }\bibfield  {title} {\bibinfo {title} {Two-dimensional magnetotransport in
  the extreme quantum limit},\ }\href
  {https://doi.org/10.1103/PhysRevLett.48.1559} {\bibfield  {journal} {\bibinfo
   {journal} {Phys. Rev. Lett.}\ }\textbf {\bibinfo {volume} {48}},\ \bibinfo
  {pages} {1559} (\bibinfo {year} {1982})}\BibitemShut {NoStop}%
\bibitem [{\citenamefont {Laughlin}(1981)}]{laughlin1981}%
  \BibitemOpen
  \bibfield  {author} {\bibinfo {author} {\bibfnamefont {R.~B.}\ \bibnamefont
  {Laughlin}},\ }\bibfield  {title} {\bibinfo {title} {Quantized hall
  conductivity in two dimensions},\ }\href
  {https://doi.org/10.1103/PhysRevB.23.5632} {\bibfield  {journal} {\bibinfo
  {journal} {Phys. Rev. B}\ }\textbf {\bibinfo {volume} {23}},\ \bibinfo
  {pages} {5632} (\bibinfo {year} {1981})}\BibitemShut {NoStop}%
\bibitem [{\citenamefont {Halperin}(1982)}]{halperin1982quantized}%
  \BibitemOpen
  \bibfield  {author} {\bibinfo {author} {\bibfnamefont {B.~I.}\ \bibnamefont
  {Halperin}},\ }\bibfield  {title} {\bibinfo {title} {Quantized hall
  conductance, current-carrying edge states, and the existence of extended
  states in a two-dimensional disordered potential},\ }\href@noop {} {\bibfield
   {journal} {\bibinfo  {journal} {Physical Review B}\ }\textbf {\bibinfo
  {volume} {25}},\ \bibinfo {pages} {2185} (\bibinfo {year}
  {1982})}\BibitemShut {NoStop}%
\bibitem [{\citenamefont {Haldane}(1983)}]{Haldane:1983xm}%
  \BibitemOpen
  \bibfield  {author} {\bibinfo {author} {\bibfnamefont {F.~D.~M.}\
  \bibnamefont {Haldane}},\ }\bibfield  {title} {\bibinfo {title} {{Fractional
  quantization of the Hall effect: A Hierarchy of incompressible quantum fluid
  states}},\ }\href {https://doi.org/10.1103/PhysRevLett.51.605} {\bibfield
  {journal} {\bibinfo  {journal} {Phys. Rev. Lett.}\ }\textbf {\bibinfo
  {volume} {51}},\ \bibinfo {pages} {605} (\bibinfo {year} {1983})}\BibitemShut
  {NoStop}%
\bibitem [{\citenamefont {Niu}\ \emph {et~al.}(1985)\citenamefont {Niu},
  \citenamefont {Thouless},\ and\ \citenamefont {Wu}}]{niu1985}%
  \BibitemOpen
  \bibfield  {author} {\bibinfo {author} {\bibfnamefont {Q.}~\bibnamefont
  {Niu}}, \bibinfo {author} {\bibfnamefont {D.~J.}\ \bibnamefont {Thouless}},\
  and\ \bibinfo {author} {\bibfnamefont {Y.-S.}\ \bibnamefont {Wu}},\
  }\bibfield  {title} {\bibinfo {title} {Quantized hall conductance as a
  topological invariant},\ }\href {https://doi.org/10.1103/PhysRevB.31.3372}
  {\bibfield  {journal} {\bibinfo  {journal} {Phys. Rev. B}\ }\textbf {\bibinfo
  {volume} {31}},\ \bibinfo {pages} {3372} (\bibinfo {year}
  {1985})}\BibitemShut {NoStop}%
\bibitem [{\citenamefont {Laughlin}(1983)}]{PhysRevLett.50.1395}%
  \BibitemOpen
  \bibfield  {author} {\bibinfo {author} {\bibfnamefont {R.~B.}\ \bibnamefont
  {Laughlin}},\ }\bibfield  {title} {\bibinfo {title} {Anomalous quantum hall
  effect: An incompressible quantum fluid with fractionally charged
  excitations},\ }\href {https://doi.org/10.1103/PhysRevLett.50.1395}
  {\bibfield  {journal} {\bibinfo  {journal} {Phys. Rev. Lett.}\ }\textbf
  {\bibinfo {volume} {50}},\ \bibinfo {pages} {1395} (\bibinfo {year}
  {1983})}\BibitemShut {NoStop}%
\bibitem [{\citenamefont {Zhang}\ \emph {et~al.}(1989)\citenamefont {Zhang},
  \citenamefont {Hansson},\ and\ \citenamefont {Kivelson}}]{PhysRevLett.62.82}%
  \BibitemOpen
  \bibfield  {author} {\bibinfo {author} {\bibfnamefont {S.~C.}\ \bibnamefont
  {Zhang}}, \bibinfo {author} {\bibfnamefont {T.~H.}\ \bibnamefont {Hansson}},\
  and\ \bibinfo {author} {\bibfnamefont {S.}~\bibnamefont {Kivelson}},\
  }\bibfield  {title} {\bibinfo {title} {Effective-field-theory model for the
  fractional quantum hall effect},\ }\href
  {https://doi.org/10.1103/PhysRevLett.62.82} {\bibfield  {journal} {\bibinfo
  {journal} {Phys. Rev. Lett.}\ }\textbf {\bibinfo {volume} {62}},\ \bibinfo
  {pages} {82} (\bibinfo {year} {1989})}\BibitemShut {NoStop}%
\bibitem [{\citenamefont {Jain}(1989)}]{PhysRevLett.63.199}%
  \BibitemOpen
  \bibfield  {author} {\bibinfo {author} {\bibfnamefont {J.~K.}\ \bibnamefont
  {Jain}},\ }\bibfield  {title} {\bibinfo {title} {Composite-fermion approach
  for the fractional quantum hall effect},\ }\href
  {https://doi.org/10.1103/PhysRevLett.63.199} {\bibfield  {journal} {\bibinfo
  {journal} {Phys. Rev. Lett.}\ }\textbf {\bibinfo {volume} {63}},\ \bibinfo
  {pages} {199} (\bibinfo {year} {1989})}\BibitemShut {NoStop}%
\bibitem [{\citenamefont {Lee}\ and\ \citenamefont
  {Kane}(1990)}]{PhysRevLett.64.1313}%
  \BibitemOpen
  \bibfield  {author} {\bibinfo {author} {\bibfnamefont {D.-H.}\ \bibnamefont
  {Lee}}\ and\ \bibinfo {author} {\bibfnamefont {C.~L.}\ \bibnamefont {Kane}},\
  }\bibfield  {title} {\bibinfo {title} {Boson-vortex-skyrmion duality,
  spin-singlet fractional quantum hall effect, and spin-1/2 anyon
  superconductivity},\ }\href {https://doi.org/10.1103/PhysRevLett.64.1313}
  {\bibfield  {journal} {\bibinfo  {journal} {Phys. Rev. Lett.}\ }\textbf
  {\bibinfo {volume} {64}},\ \bibinfo {pages} {1313} (\bibinfo {year}
  {1990})}\BibitemShut {NoStop}%
\bibitem [{\citenamefont {Kane}\ and\ \citenamefont {Mele}(2005)}]{kane2005}%
  \BibitemOpen
  \bibfield  {author} {\bibinfo {author} {\bibfnamefont {C.~L.}\ \bibnamefont
  {Kane}}\ and\ \bibinfo {author} {\bibfnamefont {E.~J.}\ \bibnamefont
  {Mele}},\ }\bibfield  {title} {\bibinfo {title} {${Z}_{2}$ topological order
  and the quantum spin hall effect},\ }\href
  {https://doi.org/10.1103/PhysRevLett.95.146802} {\bibfield  {journal}
  {\bibinfo  {journal} {Phys. Rev. Lett.}\ }\textbf {\bibinfo {volume} {95}},\
  \bibinfo {pages} {146802} (\bibinfo {year} {2005})}\BibitemShut {NoStop}%
\bibitem [{\citenamefont {Fu}\ \emph {et~al.}(2007)\citenamefont {Fu},
  \citenamefont {Kane},\ and\ \citenamefont {Mele}}]{fu_topological_2007}%
  \BibitemOpen
  \bibfield  {author} {\bibinfo {author} {\bibfnamefont {L.}~\bibnamefont
  {Fu}}, \bibinfo {author} {\bibfnamefont {C.~L.}\ \bibnamefont {Kane}},\ and\
  \bibinfo {author} {\bibfnamefont {E.~J.}\ \bibnamefont {Mele}},\ }\bibfield
  {title} {\bibinfo {title} {Topological {Insulators} in {Three}
  {Dimensions}},\ }\href {https://doi.org/10.1103/PhysRevLett.98.106803}
  {\bibfield  {journal} {\bibinfo  {journal} {Phys. Rev. Lett.}\ }\textbf
  {\bibinfo {volume} {98}},\ \bibinfo {pages} {106803} (\bibinfo {year}
  {2007})},\ \bibinfo {note} {publisher: American Physical Society}\BibitemShut
  {NoStop}%
\bibitem [{\citenamefont {Moore}\ and\ \citenamefont
  {Balents}(2007)}]{moore2007}%
  \BibitemOpen
  \bibfield  {author} {\bibinfo {author} {\bibfnamefont {J.~E.}\ \bibnamefont
  {Moore}}\ and\ \bibinfo {author} {\bibfnamefont {L.}~\bibnamefont
  {Balents}},\ }\bibfield  {title} {\bibinfo {title} {Topological invariants of
  time-reversal-invariant band structures},\ }\href
  {https://doi.org/10.1103/PhysRevB.75.121306} {\bibfield  {journal} {\bibinfo
  {journal} {Phys. Rev. B}\ }\textbf {\bibinfo {volume} {75}},\ \bibinfo
  {pages} {121306} (\bibinfo {year} {2007})}\BibitemShut {NoStop}%
\bibitem [{\citenamefont {Bernevig}\ \emph {et~al.}(2006)\citenamefont
  {Bernevig}, \citenamefont {Hughes},\ and\ \citenamefont
  {Zhang}}]{QSHI-HgTe-Theory}%
  \BibitemOpen
  \bibfield  {author} {\bibinfo {author} {\bibfnamefont {B.~A.}\ \bibnamefont
  {Bernevig}}, \bibinfo {author} {\bibfnamefont {T.~L.}\ \bibnamefont
  {Hughes}},\ and\ \bibinfo {author} {\bibfnamefont {S.-C.}\ \bibnamefont
  {Zhang}},\ }\bibfield  {title} {\bibinfo {title} {Quantum spin hall effect
  and topological phase transition in hgte quantum wells},\ }\href
  {https://doi.org/10.1126/science.1133734} {\bibfield  {journal} {\bibinfo
  {journal} {Science}\ }\textbf {\bibinfo {volume} {314}},\ \bibinfo {pages}
  {1757} (\bibinfo {year} {2006})},\ \Eprint
  {https://arxiv.org/abs/https://www.science.org/doi/pdf/10.1126/science.1133734}
  {https://www.science.org/doi/pdf/10.1126/science.1133734} \BibitemShut
  {NoStop}%
\bibitem [{\citenamefont {König}\ \emph {et~al.}(2007)\citenamefont {König},
  \citenamefont {Wiedmann}, \citenamefont {Brüne}, \citenamefont {Roth},
  \citenamefont {Buhmann}, \citenamefont {Molenkamp}, \citenamefont {Qi},\ and\
  \citenamefont {Zhang}}]{koenig2007}%
  \BibitemOpen
  \bibfield  {author} {\bibinfo {author} {\bibfnamefont {M.}~\bibnamefont
  {König}}, \bibinfo {author} {\bibfnamefont {S.}~\bibnamefont {Wiedmann}},
  \bibinfo {author} {\bibfnamefont {C.}~\bibnamefont {Brüne}}, \bibinfo
  {author} {\bibfnamefont {A.}~\bibnamefont {Roth}}, \bibinfo {author}
  {\bibfnamefont {H.}~\bibnamefont {Buhmann}}, \bibinfo {author} {\bibfnamefont
  {L.~W.}\ \bibnamefont {Molenkamp}}, \bibinfo {author} {\bibfnamefont {X.-L.}\
  \bibnamefont {Qi}},\ and\ \bibinfo {author} {\bibfnamefont {S.-C.}\
  \bibnamefont {Zhang}},\ }\bibfield  {title} {\bibinfo {title} {Quantum spin
  hall insulator state in hgte quantum wells},\ }\href
  {https://doi.org/10.1126/science.1148047} {\bibfield  {journal} {\bibinfo
  {journal} {Science}\ }\textbf {\bibinfo {volume} {318}},\ \bibinfo {pages}
  {766} (\bibinfo {year} {2007})},\ \Eprint
  {https://arxiv.org/abs/https://www.science.org/doi/pdf/10.1126/science.1148047}
  {https://www.science.org/doi/pdf/10.1126/science.1148047} \BibitemShut
  {NoStop}%
\bibitem [{\citenamefont {Hsieh}\ \emph {et~al.}(2008)\citenamefont {Hsieh},
  \citenamefont {Qian}, \citenamefont {Wray}, \citenamefont {Xia},
  \citenamefont {Hor}, \citenamefont {Cava},\ and\ \citenamefont
  {Hasan}}]{hsieh_topological_2008}%
  \BibitemOpen
  \bibfield  {author} {\bibinfo {author} {\bibfnamefont {D.}~\bibnamefont
  {Hsieh}}, \bibinfo {author} {\bibfnamefont {D.}~\bibnamefont {Qian}},
  \bibinfo {author} {\bibfnamefont {L.}~\bibnamefont {Wray}}, \bibinfo {author}
  {\bibfnamefont {Y.}~\bibnamefont {Xia}}, \bibinfo {author} {\bibfnamefont
  {Y.~S.}\ \bibnamefont {Hor}}, \bibinfo {author} {\bibfnamefont {R.~J.}\
  \bibnamefont {Cava}},\ and\ \bibinfo {author} {\bibfnamefont {M.~Z.}\
  \bibnamefont {Hasan}},\ }\bibfield  {title} {\bibinfo {title} {A topological
  {Dirac} insulator in a quantum spin {Hall} phase},\ }\href
  {https://doi.org/10.1038/nature06843} {\bibfield  {journal} {\bibinfo
  {journal} {Nature}\ }\textbf {\bibinfo {volume} {452}},\ \bibinfo {pages}
  {970} (\bibinfo {year} {2008})}\BibitemShut {NoStop}%
\bibitem [{\citenamefont {Chen}\ \emph {et~al.}(2009)\citenamefont {Chen},
  \citenamefont {Analytis}, \citenamefont {Chu}, \citenamefont {Liu},
  \citenamefont {Mo}, \citenamefont {Qi}, \citenamefont {Zhang}, \citenamefont
  {Lu}, \citenamefont {Dai}, \citenamefont {Fang}, \citenamefont {Zhang},
  \citenamefont {Fisher}, \citenamefont {Hussain},\ and\ \citenamefont
  {Shen}}]{chen2009}%
  \BibitemOpen
  \bibfield  {author} {\bibinfo {author} {\bibfnamefont {Y.~L.}\ \bibnamefont
  {Chen}}, \bibinfo {author} {\bibfnamefont {J.~G.}\ \bibnamefont {Analytis}},
  \bibinfo {author} {\bibfnamefont {J.-H.}\ \bibnamefont {Chu}}, \bibinfo
  {author} {\bibfnamefont {Z.~K.}\ \bibnamefont {Liu}}, \bibinfo {author}
  {\bibfnamefont {S.-K.}\ \bibnamefont {Mo}}, \bibinfo {author} {\bibfnamefont
  {X.~L.}\ \bibnamefont {Qi}}, \bibinfo {author} {\bibfnamefont {H.~J.}\
  \bibnamefont {Zhang}}, \bibinfo {author} {\bibfnamefont {D.~H.}\ \bibnamefont
  {Lu}}, \bibinfo {author} {\bibfnamefont {X.}~\bibnamefont {Dai}}, \bibinfo
  {author} {\bibfnamefont {Z.}~\bibnamefont {Fang}}, \bibinfo {author}
  {\bibfnamefont {S.~C.}\ \bibnamefont {Zhang}}, \bibinfo {author}
  {\bibfnamefont {I.~R.}\ \bibnamefont {Fisher}}, \bibinfo {author}
  {\bibfnamefont {Z.}~\bibnamefont {Hussain}},\ and\ \bibinfo {author}
  {\bibfnamefont {Z.-X.}\ \bibnamefont {Shen}},\ }\bibfield  {title} {\bibinfo
  {title} {Experimental realization of a three-dimensional topological
  insulator, bi<sub>2</sub>te<sub>3</sub>},\ }\href
  {https://doi.org/10.1126/science.1173034} {\bibfield  {journal} {\bibinfo
  {journal} {Science}\ }\textbf {\bibinfo {volume} {325}},\ \bibinfo {pages}
  {178} (\bibinfo {year} {2009})},\ \Eprint
  {https://arxiv.org/abs/https://www.science.org/doi/pdf/10.1126/science.1173034}
  {https://www.science.org/doi/pdf/10.1126/science.1173034} \BibitemShut
  {NoStop}%
\bibitem [{\citenamefont {Chen}\ \emph {et~al.}(2010)\citenamefont {Chen},
  \citenamefont {Chu}, \citenamefont {Analytis}, \citenamefont {Liu},
  \citenamefont {Igarashi}, \citenamefont {Kuo}, \citenamefont {Qi},
  \citenamefont {Mo}, \citenamefont {Moore}, \citenamefont {Lu}, \citenamefont
  {Hashimoto}, \citenamefont {Sasagawa}, \citenamefont {Zhang}, \citenamefont
  {Fisher}, \citenamefont {Hussain},\ and\ \citenamefont {Shen}}]{chen2010}%
  \BibitemOpen
  \bibfield  {author} {\bibinfo {author} {\bibfnamefont {Y.~L.}\ \bibnamefont
  {Chen}}, \bibinfo {author} {\bibfnamefont {J.-H.}\ \bibnamefont {Chu}},
  \bibinfo {author} {\bibfnamefont {J.~G.}\ \bibnamefont {Analytis}}, \bibinfo
  {author} {\bibfnamefont {Z.~K.}\ \bibnamefont {Liu}}, \bibinfo {author}
  {\bibfnamefont {K.}~\bibnamefont {Igarashi}}, \bibinfo {author}
  {\bibfnamefont {H.-H.}\ \bibnamefont {Kuo}}, \bibinfo {author} {\bibfnamefont
  {X.~L.}\ \bibnamefont {Qi}}, \bibinfo {author} {\bibfnamefont {S.~K.}\
  \bibnamefont {Mo}}, \bibinfo {author} {\bibfnamefont {R.~G.}\ \bibnamefont
  {Moore}}, \bibinfo {author} {\bibfnamefont {D.~H.}\ \bibnamefont {Lu}},
  \bibinfo {author} {\bibfnamefont {M.}~\bibnamefont {Hashimoto}}, \bibinfo
  {author} {\bibfnamefont {T.}~\bibnamefont {Sasagawa}}, \bibinfo {author}
  {\bibfnamefont {S.~C.}\ \bibnamefont {Zhang}}, \bibinfo {author}
  {\bibfnamefont {I.~R.}\ \bibnamefont {Fisher}}, \bibinfo {author}
  {\bibfnamefont {Z.}~\bibnamefont {Hussain}},\ and\ \bibinfo {author}
  {\bibfnamefont {Z.~X.}\ \bibnamefont {Shen}},\ }\bibfield  {title} {\bibinfo
  {title} {Massive dirac fermion on the surface of a magnetically doped
  topological insulator},\ }\href {https://doi.org/10.1126/science.1189924}
  {\bibfield  {journal} {\bibinfo  {journal} {Science}\ }\textbf {\bibinfo
  {volume} {329}},\ \bibinfo {pages} {659} (\bibinfo {year} {2010})},\ \Eprint
  {https://arxiv.org/abs/https://www.science.org/doi/pdf/10.1126/science.1189924}
  {https://www.science.org/doi/pdf/10.1126/science.1189924} \BibitemShut
  {NoStop}%
\bibitem [{\citenamefont {Peng}\ \emph {et~al.}(2010)\citenamefont {Peng},
  \citenamefont {Lai}, \citenamefont {Kong}, \citenamefont {Meister},
  \citenamefont {Chen}, \citenamefont {Qi}, \citenamefont {Zhang},
  \citenamefont {Shen},\ and\ \citenamefont {Cui}}]{peng_aharonovbohm_2010}%
  \BibitemOpen
  \bibfield  {author} {\bibinfo {author} {\bibfnamefont {H.}~\bibnamefont
  {Peng}}, \bibinfo {author} {\bibfnamefont {K.}~\bibnamefont {Lai}}, \bibinfo
  {author} {\bibfnamefont {D.}~\bibnamefont {Kong}}, \bibinfo {author}
  {\bibfnamefont {S.}~\bibnamefont {Meister}}, \bibinfo {author} {\bibfnamefont
  {Y.}~\bibnamefont {Chen}}, \bibinfo {author} {\bibfnamefont {X.-L.}\
  \bibnamefont {Qi}}, \bibinfo {author} {\bibfnamefont {S.-C.}\ \bibnamefont
  {Zhang}}, \bibinfo {author} {\bibfnamefont {Z.-X.}\ \bibnamefont {Shen}},\
  and\ \bibinfo {author} {\bibfnamefont {Y.}~\bibnamefont {Cui}},\ }\bibfield
  {title} {\bibinfo {title} {Aharonov–{Bohm} interference in topological
  insulator nanoribbons},\ }\href {https://doi.org/10.1038/nmat2609} {\bibfield
   {journal} {\bibinfo  {journal} {Nature Materials}\ }\textbf {\bibinfo
  {volume} {9}},\ \bibinfo {pages} {225} (\bibinfo {year} {2010})}\BibitemShut
  {NoStop}%
\bibitem [{\citenamefont {Liu}\ \emph {et~al.}(2014{\natexlab{a}})\citenamefont
  {Liu}, \citenamefont {Zhou}, \citenamefont {Zhang}, \citenamefont {Wang},
  \citenamefont {Weng}, \citenamefont {Prabhakaran}, \citenamefont {Mo},
  \citenamefont {Shen}, \citenamefont {Fang}, \citenamefont {Dai},
  \citenamefont {Hussain},\ and\ \citenamefont {Chen}}]{liu2014}%
  \BibitemOpen
  \bibfield  {author} {\bibinfo {author} {\bibfnamefont {Z.~K.}\ \bibnamefont
  {Liu}}, \bibinfo {author} {\bibfnamefont {B.}~\bibnamefont {Zhou}}, \bibinfo
  {author} {\bibfnamefont {Y.}~\bibnamefont {Zhang}}, \bibinfo {author}
  {\bibfnamefont {Z.~J.}\ \bibnamefont {Wang}}, \bibinfo {author}
  {\bibfnamefont {H.~M.}\ \bibnamefont {Weng}}, \bibinfo {author}
  {\bibfnamefont {D.}~\bibnamefont {Prabhakaran}}, \bibinfo {author}
  {\bibfnamefont {S.-K.}\ \bibnamefont {Mo}}, \bibinfo {author} {\bibfnamefont
  {Z.~X.}\ \bibnamefont {Shen}}, \bibinfo {author} {\bibfnamefont
  {Z.}~\bibnamefont {Fang}}, \bibinfo {author} {\bibfnamefont {X.}~\bibnamefont
  {Dai}}, \bibinfo {author} {\bibfnamefont {Z.}~\bibnamefont {Hussain}},\ and\
  \bibinfo {author} {\bibfnamefont {Y.~L.}\ \bibnamefont {Chen}},\ }\bibfield
  {title} {\bibinfo {title} {Discovery of a three-dimensional topological dirac
  semimetal, na<sub>3</sub>bi},\ }\href
  {https://doi.org/10.1126/science.1245085} {\bibfield  {journal} {\bibinfo
  {journal} {Science}\ }\textbf {\bibinfo {volume} {343}},\ \bibinfo {pages}
  {864} (\bibinfo {year} {2014}{\natexlab{a}})},\ \Eprint
  {https://arxiv.org/abs/https://www.science.org/doi/pdf/10.1126/science.1245085}
  {https://www.science.org/doi/pdf/10.1126/science.1245085} \BibitemShut
  {NoStop}%
\bibitem [{\citenamefont {Liu}\ \emph {et~al.}(2014{\natexlab{b}})\citenamefont
  {Liu}, \citenamefont {Jiang}, \citenamefont {Zhou}, \citenamefont {Wang},
  \citenamefont {Zhang}, \citenamefont {Weng}, \citenamefont {Prabhakaran},
  \citenamefont {Mo}, \citenamefont {Peng}, \citenamefont {Dudin},
  \citenamefont {Kim}, \citenamefont {Hoesch}, \citenamefont {Fang},
  \citenamefont {Dai}, \citenamefont {Shen}, \citenamefont {Feng},
  \citenamefont {Hussain},\ and\ \citenamefont {Chen}}]{liu_stable_2014}%
  \BibitemOpen
  \bibfield  {author} {\bibinfo {author} {\bibfnamefont {Z.~K.}\ \bibnamefont
  {Liu}}, \bibinfo {author} {\bibfnamefont {J.}~\bibnamefont {Jiang}}, \bibinfo
  {author} {\bibfnamefont {B.}~\bibnamefont {Zhou}}, \bibinfo {author}
  {\bibfnamefont {Z.~J.}\ \bibnamefont {Wang}}, \bibinfo {author}
  {\bibfnamefont {Y.}~\bibnamefont {Zhang}}, \bibinfo {author} {\bibfnamefont
  {H.~M.}\ \bibnamefont {Weng}}, \bibinfo {author} {\bibfnamefont
  {D.}~\bibnamefont {Prabhakaran}}, \bibinfo {author} {\bibfnamefont {S.-K.}\
  \bibnamefont {Mo}}, \bibinfo {author} {\bibfnamefont {H.}~\bibnamefont
  {Peng}}, \bibinfo {author} {\bibfnamefont {P.}~\bibnamefont {Dudin}},
  \bibinfo {author} {\bibfnamefont {T.}~\bibnamefont {Kim}}, \bibinfo {author}
  {\bibfnamefont {M.}~\bibnamefont {Hoesch}}, \bibinfo {author} {\bibfnamefont
  {Z.}~\bibnamefont {Fang}}, \bibinfo {author} {\bibfnamefont {X.}~\bibnamefont
  {Dai}}, \bibinfo {author} {\bibfnamefont {Z.~X.}\ \bibnamefont {Shen}},
  \bibinfo {author} {\bibfnamefont {D.~L.}\ \bibnamefont {Feng}}, \bibinfo
  {author} {\bibfnamefont {Z.}~\bibnamefont {Hussain}},\ and\ \bibinfo {author}
  {\bibfnamefont {Y.~L.}\ \bibnamefont {Chen}},\ }\bibfield  {title} {\bibinfo
  {title} {A stable three-dimensional topological {Dirac} semimetal {Cd3As2}},\
  }\href {https://doi.org/10.1038/nmat3990} {\bibfield  {journal} {\bibinfo
  {journal} {Nature Materials}\ }\textbf {\bibinfo {volume} {13}},\ \bibinfo
  {pages} {677} (\bibinfo {year} {2014}{\natexlab{b}})}\BibitemShut {NoStop}%
\bibitem [{\citenamefont {Xu}\ \emph {et~al.}(2015)\citenamefont {Xu},
  \citenamefont {Belopolski}, \citenamefont {Alidoust}, \citenamefont
  {Neupane}, \citenamefont {Bian}, \citenamefont {Zhang}, \citenamefont
  {Sankar}, \citenamefont {Chang}, \citenamefont {Yuan}, \citenamefont {Lee},
  \citenamefont {Huang}, \citenamefont {Zheng}, \citenamefont {Ma},
  \citenamefont {Sanchez}, \citenamefont {Wang}, \citenamefont {Bansil},
  \citenamefont {Chou}, \citenamefont {Shibayev}, \citenamefont {Lin},
  \citenamefont {Jia},\ and\ \citenamefont {Hasan}}]{xu2015}%
  \BibitemOpen
  \bibfield  {author} {\bibinfo {author} {\bibfnamefont {S.-Y.}\ \bibnamefont
  {Xu}}, \bibinfo {author} {\bibfnamefont {I.}~\bibnamefont {Belopolski}},
  \bibinfo {author} {\bibfnamefont {N.}~\bibnamefont {Alidoust}}, \bibinfo
  {author} {\bibfnamefont {M.}~\bibnamefont {Neupane}}, \bibinfo {author}
  {\bibfnamefont {G.}~\bibnamefont {Bian}}, \bibinfo {author} {\bibfnamefont
  {C.}~\bibnamefont {Zhang}}, \bibinfo {author} {\bibfnamefont
  {R.}~\bibnamefont {Sankar}}, \bibinfo {author} {\bibfnamefont
  {G.}~\bibnamefont {Chang}}, \bibinfo {author} {\bibfnamefont
  {Z.}~\bibnamefont {Yuan}}, \bibinfo {author} {\bibfnamefont {C.-C.}\
  \bibnamefont {Lee}}, \bibinfo {author} {\bibfnamefont {S.-M.}\ \bibnamefont
  {Huang}}, \bibinfo {author} {\bibfnamefont {H.}~\bibnamefont {Zheng}},
  \bibinfo {author} {\bibfnamefont {J.}~\bibnamefont {Ma}}, \bibinfo {author}
  {\bibfnamefont {D.~S.}\ \bibnamefont {Sanchez}}, \bibinfo {author}
  {\bibfnamefont {B.}~\bibnamefont {Wang}}, \bibinfo {author} {\bibfnamefont
  {A.}~\bibnamefont {Bansil}}, \bibinfo {author} {\bibfnamefont
  {F.}~\bibnamefont {Chou}}, \bibinfo {author} {\bibfnamefont {P.~P.}\
  \bibnamefont {Shibayev}}, \bibinfo {author} {\bibfnamefont {H.}~\bibnamefont
  {Lin}}, \bibinfo {author} {\bibfnamefont {S.}~\bibnamefont {Jia}},\ and\
  \bibinfo {author} {\bibfnamefont {M.~Z.}\ \bibnamefont {Hasan}},\ }\bibfield
  {title} {\bibinfo {title} {Discovery of a weyl fermion semimetal and
  topological fermi arcs},\ }\href {https://doi.org/10.1126/science.aaa9297}
  {\bibfield  {journal} {\bibinfo  {journal} {Science}\ }\textbf {\bibinfo
  {volume} {349}},\ \bibinfo {pages} {613} (\bibinfo {year} {2015})},\ \Eprint
  {https://arxiv.org/abs/https://www.science.org/doi/pdf/10.1126/science.aaa9297}
  {https://www.science.org/doi/pdf/10.1126/science.aaa9297} \BibitemShut
  {NoStop}%
\bibitem [{\citenamefont {Tang}\ \emph {et~al.}(2017)\citenamefont {Tang},
  \citenamefont {Zhang}, \citenamefont {Wong}, \citenamefont {Pedramrazi},
  \citenamefont {Tsai}, \citenamefont {Jia}, \citenamefont {Moritz},
  \citenamefont {Claassen}, \citenamefont {Ryu}, \citenamefont {Kahn},
  \citenamefont {Jiang}, \citenamefont {Yan}, \citenamefont {Hashimoto},
  \citenamefont {Lu}, \citenamefont {Moore}, \citenamefont {Hwang},
  \citenamefont {Hwang}, \citenamefont {Hussain}, \citenamefont {Chen},
  \citenamefont {Ugeda}, \citenamefont {Liu}, \citenamefont {Xie},
  \citenamefont {Devereaux}, \citenamefont {Crommie}, \citenamefont {Mo},\ and\
  \citenamefont {Shen}}]{tang_quantum_2017}%
  \BibitemOpen
  \bibfield  {author} {\bibinfo {author} {\bibfnamefont {S.}~\bibnamefont
  {Tang}}, \bibinfo {author} {\bibfnamefont {C.}~\bibnamefont {Zhang}},
  \bibinfo {author} {\bibfnamefont {D.}~\bibnamefont {Wong}}, \bibinfo {author}
  {\bibfnamefont {Z.}~\bibnamefont {Pedramrazi}}, \bibinfo {author}
  {\bibfnamefont {H.-Z.}\ \bibnamefont {Tsai}}, \bibinfo {author}
  {\bibfnamefont {C.}~\bibnamefont {Jia}}, \bibinfo {author} {\bibfnamefont
  {B.}~\bibnamefont {Moritz}}, \bibinfo {author} {\bibfnamefont
  {M.}~\bibnamefont {Claassen}}, \bibinfo {author} {\bibfnamefont
  {H.}~\bibnamefont {Ryu}}, \bibinfo {author} {\bibfnamefont {S.}~\bibnamefont
  {Kahn}}, \bibinfo {author} {\bibfnamefont {J.}~\bibnamefont {Jiang}},
  \bibinfo {author} {\bibfnamefont {H.}~\bibnamefont {Yan}}, \bibinfo {author}
  {\bibfnamefont {M.}~\bibnamefont {Hashimoto}}, \bibinfo {author}
  {\bibfnamefont {D.}~\bibnamefont {Lu}}, \bibinfo {author} {\bibfnamefont
  {R.~G.}\ \bibnamefont {Moore}}, \bibinfo {author} {\bibfnamefont {C.-C.}\
  \bibnamefont {Hwang}}, \bibinfo {author} {\bibfnamefont {C.}~\bibnamefont
  {Hwang}}, \bibinfo {author} {\bibfnamefont {Z.}~\bibnamefont {Hussain}},
  \bibinfo {author} {\bibfnamefont {Y.}~\bibnamefont {Chen}}, \bibinfo {author}
  {\bibfnamefont {M.~M.}\ \bibnamefont {Ugeda}}, \bibinfo {author}
  {\bibfnamefont {Z.}~\bibnamefont {Liu}}, \bibinfo {author} {\bibfnamefont
  {X.}~\bibnamefont {Xie}}, \bibinfo {author} {\bibfnamefont {T.~P.}\
  \bibnamefont {Devereaux}}, \bibinfo {author} {\bibfnamefont {M.~F.}\
  \bibnamefont {Crommie}}, \bibinfo {author} {\bibfnamefont {S.-K.}\
  \bibnamefont {Mo}},\ and\ \bibinfo {author} {\bibfnamefont {Z.-X.}\
  \bibnamefont {Shen}},\ }\bibfield  {title} {\bibinfo {title} {Quantum spin
  {Hall} state in monolayer {1T}'-{WTe2}},\ }\href
  {https://doi.org/10.1038/nphys4174} {\bibfield  {journal} {\bibinfo
  {journal} {Nature Physics}\ }\textbf {\bibinfo {volume} {13}},\ \bibinfo
  {pages} {683} (\bibinfo {year} {2017})}\BibitemShut {NoStop}%
\bibitem [{\citenamefont {Nadj-Perge}\ \emph {et~al.}(2014)\citenamefont
  {Nadj-Perge}, \citenamefont {Drozdov}, \citenamefont {Li}, \citenamefont
  {Chen}, \citenamefont {Jeon}, \citenamefont {Seo}, \citenamefont {MacDonald},
  \citenamefont {Bernevig},\ and\ \citenamefont {Yazdani}}]{nadjperge2014}%
  \BibitemOpen
  \bibfield  {author} {\bibinfo {author} {\bibfnamefont {S.}~\bibnamefont
  {Nadj-Perge}}, \bibinfo {author} {\bibfnamefont {I.~K.}\ \bibnamefont
  {Drozdov}}, \bibinfo {author} {\bibfnamefont {J.}~\bibnamefont {Li}},
  \bibinfo {author} {\bibfnamefont {H.}~\bibnamefont {Chen}}, \bibinfo {author}
  {\bibfnamefont {S.}~\bibnamefont {Jeon}}, \bibinfo {author} {\bibfnamefont
  {J.}~\bibnamefont {Seo}}, \bibinfo {author} {\bibfnamefont {A.~H.}\
  \bibnamefont {MacDonald}}, \bibinfo {author} {\bibfnamefont {B.~A.}\
  \bibnamefont {Bernevig}},\ and\ \bibinfo {author} {\bibfnamefont
  {A.}~\bibnamefont {Yazdani}},\ }\bibfield  {title} {\bibinfo {title}
  {Observation of majorana fermions in ferromagnetic atomic chains on a
  superconductor},\ }\href {https://doi.org/10.1126/science.1259327} {\bibfield
   {journal} {\bibinfo  {journal} {Science}\ }\textbf {\bibinfo {volume}
  {346}},\ \bibinfo {pages} {602} (\bibinfo {year} {2014})},\ \Eprint
  {https://arxiv.org/abs/https://www.science.org/doi/pdf/10.1126/science.1259327}
  {https://www.science.org/doi/pdf/10.1126/science.1259327} \BibitemShut
  {NoStop}%
\bibitem [{\citenamefont {Roushan}\ \emph {et~al.}(2009)\citenamefont
  {Roushan}, \citenamefont {Seo}, \citenamefont {Parker}, \citenamefont {Hor},
  \citenamefont {Hsieh}, \citenamefont {Qian}, \citenamefont {Richardella},
  \citenamefont {Hasan}, \citenamefont {Cava},\ and\ \citenamefont
  {Yazdani}}]{roushan_topological_2009}%
  \BibitemOpen
  \bibfield  {author} {\bibinfo {author} {\bibfnamefont {P.}~\bibnamefont
  {Roushan}}, \bibinfo {author} {\bibfnamefont {J.}~\bibnamefont {Seo}},
  \bibinfo {author} {\bibfnamefont {C.~V.}\ \bibnamefont {Parker}}, \bibinfo
  {author} {\bibfnamefont {Y.~S.}\ \bibnamefont {Hor}}, \bibinfo {author}
  {\bibfnamefont {D.}~\bibnamefont {Hsieh}}, \bibinfo {author} {\bibfnamefont
  {D.}~\bibnamefont {Qian}}, \bibinfo {author} {\bibfnamefont {A.}~\bibnamefont
  {Richardella}}, \bibinfo {author} {\bibfnamefont {M.~Z.}\ \bibnamefont
  {Hasan}}, \bibinfo {author} {\bibfnamefont {R.~J.}\ \bibnamefont {Cava}},\
  and\ \bibinfo {author} {\bibfnamefont {A.}~\bibnamefont {Yazdani}},\
  }\bibfield  {title} {\bibinfo {title} {Topological surface states protected
  from backscattering by chiral spin texture},\ }\href
  {https://doi.org/10.1038/nature08308} {\bibfield  {journal} {\bibinfo
  {journal} {Nature}\ }\textbf {\bibinfo {volume} {460}},\ \bibinfo {pages}
  {1106} (\bibinfo {year} {2009})}\BibitemShut {NoStop}%
\bibitem [{\citenamefont {Xia}\ \emph {et~al.}(2009)\citenamefont {Xia},
  \citenamefont {Qian}, \citenamefont {Hsieh}, \citenamefont {Wray},
  \citenamefont {Pal}, \citenamefont {Lin}, \citenamefont {Bansil},
  \citenamefont {Grauer}, \citenamefont {Hor}, \citenamefont {Cava},\ and\
  \citenamefont {Hasan}}]{xia_observation_2009}%
  \BibitemOpen
  \bibfield  {author} {\bibinfo {author} {\bibfnamefont {Y.}~\bibnamefont
  {Xia}}, \bibinfo {author} {\bibfnamefont {D.}~\bibnamefont {Qian}}, \bibinfo
  {author} {\bibfnamefont {D.}~\bibnamefont {Hsieh}}, \bibinfo {author}
  {\bibfnamefont {L.}~\bibnamefont {Wray}}, \bibinfo {author} {\bibfnamefont
  {A.}~\bibnamefont {Pal}}, \bibinfo {author} {\bibfnamefont {H.}~\bibnamefont
  {Lin}}, \bibinfo {author} {\bibfnamefont {A.}~\bibnamefont {Bansil}},
  \bibinfo {author} {\bibfnamefont {D.}~\bibnamefont {Grauer}}, \bibinfo
  {author} {\bibfnamefont {Y.~S.}\ \bibnamefont {Hor}}, \bibinfo {author}
  {\bibfnamefont {R.~J.}\ \bibnamefont {Cava}},\ and\ \bibinfo {author}
  {\bibfnamefont {M.~Z.}\ \bibnamefont {Hasan}},\ }\bibfield  {title} {\bibinfo
  {title} {Observation of a large-gap topological-insulator class with a single
  {Dirac} cone on the surface},\ }\href {https://doi.org/10.1038/nphys1274}
  {\bibfield  {journal} {\bibinfo  {journal} {Nature Physics}\ }\textbf
  {\bibinfo {volume} {5}},\ \bibinfo {pages} {398} (\bibinfo {year}
  {2009})}\BibitemShut {NoStop}%
\bibitem [{\citenamefont {Huang}\ \emph {et~al.}(2015)\citenamefont {Huang},
  \citenamefont {Xu}, \citenamefont {Belopolski}, \citenamefont {Lee},
  \citenamefont {Chang}, \citenamefont {Wang}, \citenamefont {Alidoust},
  \citenamefont {Bian}, \citenamefont {Neupane}, \citenamefont {Zhang},
  \citenamefont {Jia}, \citenamefont {Bansil}, \citenamefont {Lin},\ and\
  \citenamefont {Hasan}}]{huang_weyl_2015}%
  \BibitemOpen
  \bibfield  {author} {\bibinfo {author} {\bibfnamefont {S.-M.}\ \bibnamefont
  {Huang}}, \bibinfo {author} {\bibfnamefont {S.-Y.}\ \bibnamefont {Xu}},
  \bibinfo {author} {\bibfnamefont {I.}~\bibnamefont {Belopolski}}, \bibinfo
  {author} {\bibfnamefont {C.-C.}\ \bibnamefont {Lee}}, \bibinfo {author}
  {\bibfnamefont {G.}~\bibnamefont {Chang}}, \bibinfo {author} {\bibfnamefont
  {B.}~\bibnamefont {Wang}}, \bibinfo {author} {\bibfnamefont {N.}~\bibnamefont
  {Alidoust}}, \bibinfo {author} {\bibfnamefont {G.}~\bibnamefont {Bian}},
  \bibinfo {author} {\bibfnamefont {M.}~\bibnamefont {Neupane}}, \bibinfo
  {author} {\bibfnamefont {C.}~\bibnamefont {Zhang}}, \bibinfo {author}
  {\bibfnamefont {S.}~\bibnamefont {Jia}}, \bibinfo {author} {\bibfnamefont
  {A.}~\bibnamefont {Bansil}}, \bibinfo {author} {\bibfnamefont
  {H.}~\bibnamefont {Lin}},\ and\ \bibinfo {author} {\bibfnamefont {M.~Z.}\
  \bibnamefont {Hasan}},\ }\bibfield  {title} {\bibinfo {title} {A {Weyl}
  {Fermion} semimetal with surface {Fermi} arcs in the transition metal
  monopnictide {TaAs} class},\ }\href {https://doi.org/10.1038/ncomms8373}
  {\bibfield  {journal} {\bibinfo  {journal} {Nature Communications}\ }\textbf
  {\bibinfo {volume} {6}},\ \bibinfo {pages} {7373} (\bibinfo {year}
  {2015})}\BibitemShut {NoStop}%
\bibitem [{\citenamefont {Kitaev}(2003)}]{KITAEV20032}%
  \BibitemOpen
  \bibfield  {author} {\bibinfo {author} {\bibfnamefont {A.}~\bibnamefont
  {Kitaev}},\ }\bibfield  {title} {\bibinfo {title} {Fault-tolerant quantum
  computation by anyons},\ }\href
  {https://doi.org/https://doi.org/10.1016/S0003-4916(02)00018-0} {\bibfield
  {journal} {\bibinfo  {journal} {Annals of Physics}\ }\textbf {\bibinfo
  {volume} {303}},\ \bibinfo {pages} {2} (\bibinfo {year} {2003})}\BibitemShut
  {NoStop}%
\bibitem [{\citenamefont {Kitaev}(2006)}]{KITAEV20062}%
  \BibitemOpen
  \bibfield  {author} {\bibinfo {author} {\bibfnamefont {A.}~\bibnamefont
  {Kitaev}},\ }\bibfield  {title} {\bibinfo {title} {Anyons in an exactly
  solved model and beyond},\ }\href
  {https://doi.org/https://doi.org/10.1016/j.aop.2005.10.005} {\bibfield
  {journal} {\bibinfo  {journal} {Annals of Physics}\ }\textbf {\bibinfo
  {volume} {321}},\ \bibinfo {pages} {2} (\bibinfo {year} {2006})},\ \bibinfo
  {note} {january Special Issue}\BibitemShut {NoStop}%
\bibitem [{\citenamefont {Read}\ and\ \citenamefont {Green}(2000)}]{read2000}%
  \BibitemOpen
  \bibfield  {author} {\bibinfo {author} {\bibfnamefont {N.}~\bibnamefont
  {Read}}\ and\ \bibinfo {author} {\bibfnamefont {D.}~\bibnamefont {Green}},\
  }\bibfield  {title} {\bibinfo {title} {Paired states of fermions in two
  dimensions with breaking of parity and time-reversal symmetries and the
  fractional quantum hall effect},\ }\href
  {https://doi.org/10.1103/PhysRevB.61.10267} {\bibfield  {journal} {\bibinfo
  {journal} {Phys. Rev. B}\ }\textbf {\bibinfo {volume} {61}},\ \bibinfo
  {pages} {10267} (\bibinfo {year} {2000})}\BibitemShut {NoStop}%
\bibitem [{\citenamefont {Nayak}\ \emph {et~al.}(2008)\citenamefont {Nayak},
  \citenamefont {Simon}, \citenamefont {Stern}, \citenamefont {Freedman},\ and\
  \citenamefont {Das~Sarma}}]{nayak2008}%
  \BibitemOpen
  \bibfield  {author} {\bibinfo {author} {\bibfnamefont {C.}~\bibnamefont
  {Nayak}}, \bibinfo {author} {\bibfnamefont {S.~H.}\ \bibnamefont {Simon}},
  \bibinfo {author} {\bibfnamefont {A.}~\bibnamefont {Stern}}, \bibinfo
  {author} {\bibfnamefont {M.}~\bibnamefont {Freedman}},\ and\ \bibinfo
  {author} {\bibfnamefont {S.}~\bibnamefont {Das~Sarma}},\ }\bibfield  {title}
  {\bibinfo {title} {Non-abelian anyons and topological quantum computation},\
  }\href {https://doi.org/10.1103/RevModPhys.80.1083} {\bibfield  {journal}
  {\bibinfo  {journal} {Rev. Mod. Phys.}\ }\textbf {\bibinfo {volume} {80}},\
  \bibinfo {pages} {1083} (\bibinfo {year} {2008})}\BibitemShut {NoStop}%
\bibitem [{\citenamefont {Ryu}\ \emph {et~al.}(2010)\citenamefont {Ryu},
  \citenamefont {Schnyder}, \citenamefont {Furusaki},\ and\ \citenamefont
  {Ludwig}}]{Ryu_2010}%
  \BibitemOpen
  \bibfield  {author} {\bibinfo {author} {\bibfnamefont {S.}~\bibnamefont
  {Ryu}}, \bibinfo {author} {\bibfnamefont {A.~P.}\ \bibnamefont {Schnyder}},
  \bibinfo {author} {\bibfnamefont {A.}~\bibnamefont {Furusaki}},\ and\
  \bibinfo {author} {\bibfnamefont {A.~W.~W.}\ \bibnamefont {Ludwig}},\
  }\bibfield  {title} {\bibinfo {title} {Topological insulators and
  superconductors: tenfold way and dimensional hierarchy},\ }\href
  {https://doi.org/10.1088/1367-2630/12/6/065010} {\bibfield  {journal}
  {\bibinfo  {journal} {New Journal of Physics}\ }\textbf {\bibinfo {volume}
  {12}},\ \bibinfo {pages} {065010} (\bibinfo {year} {2010})}\BibitemShut
  {NoStop}%
\bibitem [{\citenamefont {Schnyder}\ \emph {et~al.}(2008)\citenamefont
  {Schnyder}, \citenamefont {Ryu}, \citenamefont {Furusaki},\ and\
  \citenamefont {Ludwig}}]{schnyder2008}%
  \BibitemOpen
  \bibfield  {author} {\bibinfo {author} {\bibfnamefont {A.~P.}\ \bibnamefont
  {Schnyder}}, \bibinfo {author} {\bibfnamefont {S.}~\bibnamefont {Ryu}},
  \bibinfo {author} {\bibfnamefont {A.}~\bibnamefont {Furusaki}},\ and\
  \bibinfo {author} {\bibfnamefont {A.~W.~W.}\ \bibnamefont {Ludwig}},\
  }\bibfield  {title} {\bibinfo {title} {Classification of topological
  insulators and superconductors in three spatial dimensions},\ }\href
  {https://doi.org/10.1103/PhysRevB.78.195125} {\bibfield  {journal} {\bibinfo
  {journal} {Phys. Rev. B}\ }\textbf {\bibinfo {volume} {78}},\ \bibinfo
  {pages} {195125} (\bibinfo {year} {2008})}\BibitemShut {NoStop}%
\bibitem [{\citenamefont {Kitaev}(2009)}]{kitaev_periodic_2009}%
  \BibitemOpen
  \bibfield  {author} {\bibinfo {author} {\bibfnamefont {A.}~\bibnamefont
  {Kitaev}},\ }\bibfield  {title} {\bibinfo {title} {Periodic table for
  topological insulators and superconductors},\ }\href
  {https://doi.org/10.1063/1.3149495} {\bibfield  {journal} {\bibinfo
  {journal} {AIP Conference Proceedings}\ }\textbf {\bibinfo {volume} {1134}},\
  \bibinfo {pages} {22} (\bibinfo {year} {2009})},\ \bibinfo {note} {\_eprint:
  https://pubs.aip.org/aip/acp/article-pdf/1134/1/22/11584243/22\_1\_online.pdf}\BibitemShut
  {NoStop}%
\bibitem [{\citenamefont {Cook}(2024)}]{qskhe}%
  \BibitemOpen
  \bibfield  {author} {\bibinfo {author} {\bibfnamefont {A.~M.}\ \bibnamefont
  {Cook}},\ }\bibfield  {title} {\bibinfo {title} {Quantum skyrmion hall
  effect},\ }\href {https://doi.org/10.1103/PhysRevB.109.155123} {\bibfield
  {journal} {\bibinfo  {journal} {Phys. Rev. B}\ }\textbf {\bibinfo {volume}
  {109}},\ \bibinfo {pages} {155123} (\bibinfo {year} {2024})}\BibitemShut
  {NoStop}%
\bibitem [{\citenamefont {Patil}\ \emph {et~al.}(2024)\citenamefont {Patil},
  \citenamefont {Flores-Caldéron},\ and\ \citenamefont {Cook}}]{patil2024}%
  \BibitemOpen
  \bibfield  {author} {\bibinfo {author} {\bibfnamefont {V.}~\bibnamefont
  {Patil}}, \bibinfo {author} {\bibfnamefont {R.}~\bibnamefont
  {Flores-Caldéron}},\ and\ \bibinfo {author} {\bibfnamefont {A.~M.}\
  \bibnamefont {Cook}},\ }\href@noop {} {\bibinfo {title} {Effective field
  theory of the quantum skyrmion hall effect}} (\bibinfo {year} {2024}),\
  \Eprint {https://arxiv.org/abs/blah} {arXiv:blah [quant-ph]} \BibitemShut
  {NoStop}%
\bibitem [{\citenamefont {Bernevig}\ \emph {et~al.}(2002)\citenamefont
  {Bernevig}, \citenamefont {Chern}, \citenamefont {Hu}, \citenamefont
  {Toumbas},\ and\ \citenamefont {Zhang}}]{bernevig6Dfieldtheory}%
  \BibitemOpen
  \bibfield  {author} {\bibinfo {author} {\bibfnamefont {B.~A.}\ \bibnamefont
  {Bernevig}}, \bibinfo {author} {\bibfnamefont {C.-H.}\ \bibnamefont {Chern}},
  \bibinfo {author} {\bibfnamefont {J.-P.}\ \bibnamefont {Hu}}, \bibinfo
  {author} {\bibfnamefont {N.}~\bibnamefont {Toumbas}},\ and\ \bibinfo {author}
  {\bibfnamefont {S.-C.}\ \bibnamefont {Zhang}},\ }\bibfield  {title} {\bibinfo
  {title} {Effective field theory description of the higher dimensional quantum
  hall liquid},\ }\href
  {https://doi.org/https://doi.org/10.1006/aphy.2002.6292} {\bibfield
  {journal} {\bibinfo  {journal} {Annals of Physics}\ }\textbf {\bibinfo
  {volume} {300}},\ \bibinfo {pages} {185} (\bibinfo {year}
  {2002})}\BibitemShut {NoStop}%
\bibitem [{\citenamefont {Cook}\ and\ \citenamefont
  {Moore}(2022)}]{cook_multiplicative_2022}%
  \BibitemOpen
  \bibfield  {author} {\bibinfo {author} {\bibfnamefont {A.~M.}\ \bibnamefont
  {Cook}}\ and\ \bibinfo {author} {\bibfnamefont {J.~E.}\ \bibnamefont
  {Moore}},\ }\bibfield  {title} {\bibinfo {title} {Multiplicative topological
  phases},\ }\href {https://doi.org/10.1038/s42005-022-01022-x} {\bibfield
  {journal} {\bibinfo  {journal} {Communications Physics}\ }\textbf {\bibinfo
  {volume} {5}},\ \bibinfo {pages} {262} (\bibinfo {year} {2022})}\BibitemShut
  {NoStop}%
\bibitem [{\citenamefont {Cook}(2023)}]{cook2023}%
  \BibitemOpen
  \bibfield  {author} {\bibinfo {author} {\bibfnamefont {A.~M.}\ \bibnamefont
  {Cook}},\ }\bibfield  {title} {\bibinfo {title} {Topological skyrmion phases
  of matter},\ }\href
  {http://iopscience.iop.org/article/10.1088/1361-648X/acbffd} {\bibfield
  {journal} {\bibinfo  {journal} {Journal of Physics: Condensed Matter}\
  }\textbf {\bibinfo {volume} {35}},\ \bibinfo {pages} {184001} (\bibinfo
  {year} {2023})}\BibitemShut {NoStop}%
\bibitem [{\citenamefont {Cook}\ and\ \citenamefont
  {Nielsen}(2023)}]{cookFST2023}%
  \BibitemOpen
  \bibfield  {author} {\bibinfo {author} {\bibfnamefont {A.~M.}\ \bibnamefont
  {Cook}}\ and\ \bibinfo {author} {\bibfnamefont {A.~E.~B.}\ \bibnamefont
  {Nielsen}},\ }\bibfield  {title} {\bibinfo {title} {Finite-size topology},\
  }\href {https://doi.org/10.1103/PhysRevB.108.045144} {\bibfield  {journal}
  {\bibinfo  {journal} {Phys. Rev. B}\ }\textbf {\bibinfo {volume} {108}},\
  \bibinfo {pages} {045144} (\bibinfo {year} {2023})}\BibitemShut {NoStop}%
\bibitem [{\citenamefont {Liu}\ \emph {et~al.}(2023{\natexlab{a}})\citenamefont
  {Liu}, \citenamefont {Shi},\ and\ \citenamefont {Cook}}]{liu2020}%
  \BibitemOpen
  \bibfield  {author} {\bibinfo {author} {\bibfnamefont {S.-W.}\ \bibnamefont
  {Liu}}, \bibinfo {author} {\bibfnamefont {L.-k.}\ \bibnamefont {Shi}},\ and\
  \bibinfo {author} {\bibfnamefont {A.~M.}\ \bibnamefont {Cook}},\ }\bibfield
  {title} {\bibinfo {title} {Defect bulk-boundary correspondence of topological
  skyrmion phases of matter},\ }\href
  {https://doi.org/10.1103/PhysRevB.107.235109} {\bibfield  {journal} {\bibinfo
   {journal} {Phys. Rev. B}\ }\textbf {\bibinfo {volume} {107}},\ \bibinfo
  {pages} {235109} (\bibinfo {year} {2023}{\natexlab{a}})}\BibitemShut
  {NoStop}%
\bibitem [{\citenamefont {Pal}\ \emph {et~al.}(2024)\citenamefont {Pal},
  \citenamefont {Winter},\ and\ \citenamefont {Cook}}]{pal_multsemimetal}%
  \BibitemOpen
  \bibfield  {author} {\bibinfo {author} {\bibfnamefont {A.}~\bibnamefont
  {Pal}}, \bibinfo {author} {\bibfnamefont {J.~H.}\ \bibnamefont {Winter}},\
  and\ \bibinfo {author} {\bibfnamefont {A.~M.}\ \bibnamefont {Cook}},\
  }\bibfield  {title} {\bibinfo {title} {Multiplicative topological
  semimetals},\ }\href {https://doi.org/10.1103/PhysRevB.109.035147} {\bibfield
   {journal} {\bibinfo  {journal} {Phys. Rev. B}\ }\textbf {\bibinfo {volume}
  {109}},\ \bibinfo {pages} {035147} (\bibinfo {year} {2024})}\BibitemShut
  {NoStop}%
\bibitem [{\citenamefont {Pal}\ and\ \citenamefont
  {Cook}(2024)}]{pal2024_fstwsm}%
  \BibitemOpen
  \bibfield  {author} {\bibinfo {author} {\bibfnamefont {A.}~\bibnamefont
  {Pal}}\ and\ \bibinfo {author} {\bibfnamefont {A.~M.}\ \bibnamefont {Cook}},\
  }\href {https://arxiv.org/abs/2409.05842} {\bibinfo {title} {Finite-size
  topological phases from semimetals}} (\bibinfo {year} {2024}),\ \Eprint
  {https://arxiv.org/abs/2409.05842} {arXiv:2409.05842 [cond-mat.mes-hall]}
  \BibitemShut {NoStop}%
\bibitem [{\citenamefont {Winter}\ \emph {et~al.}(2023)\citenamefont {Winter},
  \citenamefont {Ay}, \citenamefont {Braunecker},\ and\ \citenamefont
  {Cook}}]{winterOEPT}%
  \BibitemOpen
  \bibfield  {author} {\bibinfo {author} {\bibfnamefont {J.~H.}\ \bibnamefont
  {Winter}}, \bibinfo {author} {\bibfnamefont {R.}~\bibnamefont {Ay}}, \bibinfo
  {author} {\bibfnamefont {B.}~\bibnamefont {Braunecker}},\ and\ \bibinfo
  {author} {\bibfnamefont {A.~M.}\ \bibnamefont {Cook}},\ }\href@noop {}
  {\bibinfo {title} {Observable-enriched entanglement}} (\bibinfo {year}
  {2023}),\ \Eprint {https://arxiv.org/abs/2312.09153} {arXiv:2312.09153
  [quant-ph]} \BibitemShut {NoStop}%
\bibitem [{\citenamefont {Flores-Calderon}\ and\ \citenamefont
  {Cook}(2023)}]{calderon_skyrm}%
  \BibitemOpen
  \bibfield  {author} {\bibinfo {author} {\bibfnamefont {R.}~\bibnamefont
  {Flores-Calderon}}\ and\ \bibinfo {author} {\bibfnamefont {A.~M.}\
  \bibnamefont {Cook}},\ }\bibfield  {title} {\bibinfo {title} {Time-reversal
  invariant topological skyrmion phases},\ }\href
  {https://doi.org/10.1103/PhysRevB.108.235102} {\bibfield  {journal} {\bibinfo
   {journal} {Phys. Rev. B}\ }\textbf {\bibinfo {volume} {108}},\ \bibinfo
  {pages} {235102} (\bibinfo {year} {2023})}\BibitemShut {NoStop}%
\bibitem [{\citenamefont {Flores-Calderon}\ \emph {et~al.}(2023)\citenamefont
  {Flores-Calderon}, \citenamefont {Moessner},\ and\ \citenamefont
  {Cook}}]{calderon2023_fst}%
  \BibitemOpen
  \bibfield  {author} {\bibinfo {author} {\bibfnamefont {R.}~\bibnamefont
  {Flores-Calderon}}, \bibinfo {author} {\bibfnamefont {R.}~\bibnamefont
  {Moessner}},\ and\ \bibinfo {author} {\bibfnamefont {A.~M.}\ \bibnamefont
  {Cook}},\ }\bibfield  {title} {\bibinfo {title} {Time-reversal invariant
  finite-size topology},\ }\href {https://doi.org/10.1103/PhysRevB.108.125410}
  {\bibfield  {journal} {\bibinfo  {journal} {Phys. Rev. B}\ }\textbf {\bibinfo
  {volume} {108}},\ \bibinfo {pages} {125410} (\bibinfo {year}
  {2023})}\BibitemShut {NoStop}%
\bibitem [{\citenamefont {Flores-Calderón}\ \emph {et~al.}(2023)\citenamefont
  {Flores-Calderón}, \citenamefont {König},\ and\ \citenamefont
  {Cook}}]{calderongapless2023}%
  \BibitemOpen
  \bibfield  {author} {\bibinfo {author} {\bibfnamefont {R.}~\bibnamefont
  {Flores-Calderón}}, \bibinfo {author} {\bibfnamefont {E.~J.}\ \bibnamefont
  {König}},\ and\ \bibinfo {author} {\bibfnamefont {A.~M.}\ \bibnamefont
  {Cook}},\ }\href {https://arxiv.org/abs/2311.17799} {\bibinfo {title}
  {Topological quantum criticality from multiplicative topological phases}}
  (\bibinfo {year} {2023}),\ \Eprint {https://arxiv.org/abs/2311.17799}
  {arXiv:2311.17799 [cond-mat.str-el]} \BibitemShut {NoStop}%
\bibitem [{\citenamefont {Ay}\ \emph {et~al.}(2023)\citenamefont {Ay},
  \citenamefont {Winter},\ and\ \citenamefont {Cook}}]{ay2023}%
  \BibitemOpen
  \bibfield  {author} {\bibinfo {author} {\bibfnamefont {R.}~\bibnamefont
  {Ay}}, \bibinfo {author} {\bibfnamefont {J.~H.}\ \bibnamefont {Winter}},\
  and\ \bibinfo {author} {\bibfnamefont {A.~M.}\ \bibnamefont {Cook}},\ }\href
  {https://arxiv.org/abs/2311.15694} {\bibinfo {title} {Type-ii topological
  phase transitions of topological skyrmion phases}} (\bibinfo {year} {2023}),\
  \Eprint {https://arxiv.org/abs/2311.15694} {arXiv:2311.15694
  [cond-mat.mes-hall]} \BibitemShut {NoStop}%
\bibitem [{\citenamefont {Pacholski}\ and\ \citenamefont
  {Cook}(2024)}]{pacholski2024}%
  \BibitemOpen
  \bibfield  {author} {\bibinfo {author} {\bibfnamefont {M.~J.}\ \bibnamefont
  {Pacholski}}\ and\ \bibinfo {author} {\bibfnamefont {A.~M.}\ \bibnamefont
  {Cook}},\ }\bibfield  {title} {\bibinfo {title} {Crystalline finite-size
  topology},\ }\href {https://doi.org/10.1103/PhysRevB.109.235125} {\bibfield
  {journal} {\bibinfo  {journal} {Phys. Rev. B}\ }\textbf {\bibinfo {volume}
  {109}},\ \bibinfo {pages} {235125} (\bibinfo {year} {2024})}\BibitemShut
  {NoStop}%
\bibitem [{\citenamefont {Liu}\ \emph {et~al.}(2023{\natexlab{b}})\citenamefont
  {Liu}, \citenamefont {Winter},\ and\ \citenamefont
  {Cook}}]{liu2023skyrmsemimetals}%
  \BibitemOpen
  \bibfield  {author} {\bibinfo {author} {\bibfnamefont {S.-W.}\ \bibnamefont
  {Liu}}, \bibinfo {author} {\bibfnamefont {J.~H.}\ \bibnamefont {Winter}},\
  and\ \bibinfo {author} {\bibfnamefont {A.~M.}\ \bibnamefont {Cook}},\ }\href
  {https://arxiv.org/abs/2311.15753} {\bibinfo {title} {Topological skyrmion
  semimetals}} (\bibinfo {year} {2023}{\natexlab{b}}),\ \Eprint
  {https://arxiv.org/abs/2311.15753} {arXiv:2311.15753 [cond-mat.mes-hall]}
  \BibitemShut {NoStop}%
\bibitem [{\citenamefont {Qi}\ \emph {et~al.}(2008)\citenamefont {Qi},
  \citenamefont {Hughes},\ and\ \citenamefont {Zhang}}]{qi2008TRIFT}%
  \BibitemOpen
  \bibfield  {author} {\bibinfo {author} {\bibfnamefont {X.-L.}\ \bibnamefont
  {Qi}}, \bibinfo {author} {\bibfnamefont {T.~L.}\ \bibnamefont {Hughes}},\
  and\ \bibinfo {author} {\bibfnamefont {S.-C.}\ \bibnamefont {Zhang}},\
  }\bibfield  {title} {\bibinfo {title} {Topological field theory of
  time-reversal invariant insulators},\ }\href
  {https://doi.org/10.1103/PhysRevB.78.195424} {\bibfield  {journal} {\bibinfo
  {journal} {Phys. Rev. B}\ }\textbf {\bibinfo {volume} {78}},\ \bibinfo
  {pages} {195424} (\bibinfo {year} {2008})}\BibitemShut {NoStop}%
\bibitem [{\citenamefont {Aschieri}\ \emph {et~al.}(2007)\citenamefont
  {Aschieri}, \citenamefont {Steinacker}, \citenamefont {Madore}, \citenamefont
  {Manousselis},\ and\ \citenamefont {Zoupanos}}]{aschieri2007}%
  \BibitemOpen
  \bibfield  {author} {\bibinfo {author} {\bibfnamefont {P.}~\bibnamefont
  {Aschieri}}, \bibinfo {author} {\bibfnamefont {H.}~\bibnamefont
  {Steinacker}}, \bibinfo {author} {\bibfnamefont {J.}~\bibnamefont {Madore}},
  \bibinfo {author} {\bibfnamefont {P.}~\bibnamefont {Manousselis}},\ and\
  \bibinfo {author} {\bibfnamefont {G.}~\bibnamefont {Zoupanos}},\ }\bibfield
  {title} {\bibinfo {title} {Fuzzy extra dimensions: Dimensional reduction,
  dynamical generation and renormalizability},\ }\href@noop {} {\bibfield
  {journal} {\bibinfo  {journal} {Proceedings of the 4th Summer School in
  Modern Mathematical Physics, MPHYS 2006}\ } (\bibinfo {year}
  {2007})}\BibitemShut {NoStop}%
\bibitem [{\citenamefont {Aschieri}\ \emph
  {et~al.}(2004{\natexlab{a}})\citenamefont {Aschieri}, \citenamefont {Madore},
  \citenamefont {Manousselis},\ and\ \citenamefont
  {Zoupanos}}]{Aschieri:2003vy}%
  \BibitemOpen
  \bibfield  {author} {\bibinfo {author} {\bibfnamefont {P.}~\bibnamefont
  {Aschieri}}, \bibinfo {author} {\bibfnamefont {J.}~\bibnamefont {Madore}},
  \bibinfo {author} {\bibfnamefont {P.}~\bibnamefont {Manousselis}},\ and\
  \bibinfo {author} {\bibfnamefont {G.}~\bibnamefont {Zoupanos}},\ }\bibfield
  {title} {\bibinfo {title} {{Dimensional reduction over fuzzy coset spaces}},\
  }\href {https://doi.org/10.1088/1126-6708/2004/04/034} {\bibfield  {journal}
  {\bibinfo  {journal} {JHEP}\ }\textbf {\bibinfo {volume} {04}},\ \bibinfo
  {pages} {034}},\ \Eprint {https://arxiv.org/abs/hep-th/0310072}
  {arXiv:hep-th/0310072} \BibitemShut {NoStop}%
\bibitem [{\citenamefont {Aschieri}\ \emph
  {et~al.}(2004{\natexlab{b}})\citenamefont {Aschieri}, \citenamefont {Madore},
  \citenamefont {Manousselis},\ and\ \citenamefont
  {Zoupanos}}]{Aschieri:2004vh}%
  \BibitemOpen
  \bibfield  {author} {\bibinfo {author} {\bibfnamefont {P.}~\bibnamefont
  {Aschieri}}, \bibinfo {author} {\bibfnamefont {J.}~\bibnamefont {Madore}},
  \bibinfo {author} {\bibfnamefont {P.}~\bibnamefont {Manousselis}},\ and\
  \bibinfo {author} {\bibfnamefont {G.}~\bibnamefont {Zoupanos}},\ }\bibfield
  {title} {\bibinfo {title} {{Unified theories from fuzzy extra dimensions}},\
  }\href {https://doi.org/10.1002/prop.200410168} {\bibfield  {journal}
  {\bibinfo  {journal} {Fortsch. Phys.}\ }\textbf {\bibinfo {volume} {52}},\
  \bibinfo {pages} {718} (\bibinfo {year} {2004}{\natexlab{b}})},\ \Eprint
  {https://arxiv.org/abs/hep-th/0401200} {arXiv:hep-th/0401200} \BibitemShut
  {NoStop}%
\bibitem [{\citenamefont {Zhang}\ and\ \citenamefont {Hu}(2001)}]{zhanghu2001}%
  \BibitemOpen
  \bibfield  {author} {\bibinfo {author} {\bibfnamefont {S.-C.}\ \bibnamefont
  {Zhang}}\ and\ \bibinfo {author} {\bibfnamefont {J.}~\bibnamefont {Hu}},\
  }\bibfield  {title} {\bibinfo {title} {A four-dimensional generalization of
  the quantum hall effect},\ }\href
  {https://doi.org/10.1126/science.294.5543.823} {\bibfield  {journal}
  {\bibinfo  {journal} {Science}\ }\textbf {\bibinfo {volume} {294}},\ \bibinfo
  {pages} {823} (\bibinfo {year} {2001})},\ \Eprint
  {https://arxiv.org/abs/https://www.science.org/doi/pdf/10.1126/science.294.5543.823}
  {https://www.science.org/doi/pdf/10.1126/science.294.5543.823} \BibitemShut
  {NoStop}%
\bibitem [{\citenamefont {Qi}\ \emph {et~al.}(2006)\citenamefont {Qi},
  \citenamefont {Wu},\ and\ \citenamefont {Zhang}}]{qi2006_QWZmodel}%
  \BibitemOpen
  \bibfield  {author} {\bibinfo {author} {\bibfnamefont {X.-L.}\ \bibnamefont
  {Qi}}, \bibinfo {author} {\bibfnamefont {Y.-S.}\ \bibnamefont {Wu}},\ and\
  \bibinfo {author} {\bibfnamefont {S.-C.}\ \bibnamefont {Zhang}},\ }\bibfield
  {title} {\bibinfo {title} {Topological quantization of the spin {H}all effect
  in two-dimensional paramagnetic semiconductors},\ }\href
  {https://doi.org/10.1103/PhysRevB.74.085308} {\bibfield  {journal} {\bibinfo
  {journal} {Phys. Rev. B}\ }\textbf {\bibinfo {volume} {74}},\ \bibinfo
  {pages} {085308} (\bibinfo {year} {2006})}\BibitemShut {NoStop}%
\bibitem [{\citenamefont {Nielsen}\ and\ \citenamefont
  {Ninomiya}(1981)}]{Nielsen:1981hk}%
  \BibitemOpen
  \bibfield  {author} {\bibinfo {author} {\bibfnamefont {H.~B.}\ \bibnamefont
  {Nielsen}}\ and\ \bibinfo {author} {\bibfnamefont {M.}~\bibnamefont
  {Ninomiya}},\ }\bibfield  {title} {\bibinfo {title} {{No Go Theorem for
  Regularizing Chiral Fermions}},\ }\href
  {https://doi.org/10.1016/0370-2693(81)91026-1} {\bibfield  {journal}
  {\bibinfo  {journal} {Phys. Lett. B}\ }\textbf {\bibinfo {volume} {105}},\
  \bibinfo {pages} {219} (\bibinfo {year} {1981})}\BibitemShut {NoStop}%
\bibitem [{\citenamefont {Peschel}(2003)}]{IngoPeschel_2003}%
  \BibitemOpen
  \bibfield  {author} {\bibinfo {author} {\bibfnamefont {I.}~\bibnamefont
  {Peschel}},\ }\bibfield  {title} {\bibinfo {title} {Calculation of reduced
  density matrices from correlation functions},\ }\href
  {https://doi.org/10.1088/0305-4470/36/14/101} {\bibfield  {journal} {\bibinfo
   {journal} {Journal of Physics A: Mathematical and General}\ }\textbf
  {\bibinfo {volume} {36}},\ \bibinfo {pages} {L205} (\bibinfo {year}
  {2003})}\BibitemShut {NoStop}%
\bibitem [{\citenamefont {Ma}\ \emph {et~al.}(2015)\citenamefont {Ma},
  \citenamefont {Calvo}, \citenamefont {Wang}, \citenamefont {Lian},
  \citenamefont {Mühlbauer}, \citenamefont {Brüne}, \citenamefont {Cui},
  \citenamefont {Lai}, \citenamefont {Kundhikanjana}, \citenamefont {Yang},
  \citenamefont {Baenninger}, \citenamefont {König}, \citenamefont {Ames},
  \citenamefont {Buhmann}, \citenamefont {Leubner}, \citenamefont {Molenkamp},
  \citenamefont {Zhang}, \citenamefont {Goldhaber-Gordon}, \citenamefont
  {Kelly},\ and\ \citenamefont {Shen}}]{ma_unexpected_2015}%
  \BibitemOpen
  \bibfield  {author} {\bibinfo {author} {\bibfnamefont {E.~Y.}\ \bibnamefont
  {Ma}}, \bibinfo {author} {\bibfnamefont {M.~R.}\ \bibnamefont {Calvo}},
  \bibinfo {author} {\bibfnamefont {J.}~\bibnamefont {Wang}}, \bibinfo {author}
  {\bibfnamefont {B.}~\bibnamefont {Lian}}, \bibinfo {author} {\bibfnamefont
  {M.}~\bibnamefont {Mühlbauer}}, \bibinfo {author} {\bibfnamefont
  {C.}~\bibnamefont {Brüne}}, \bibinfo {author} {\bibfnamefont {Y.-T.}\
  \bibnamefont {Cui}}, \bibinfo {author} {\bibfnamefont {K.}~\bibnamefont
  {Lai}}, \bibinfo {author} {\bibfnamefont {W.}~\bibnamefont {Kundhikanjana}},
  \bibinfo {author} {\bibfnamefont {Y.}~\bibnamefont {Yang}}, \bibinfo {author}
  {\bibfnamefont {M.}~\bibnamefont {Baenninger}}, \bibinfo {author}
  {\bibfnamefont {M.}~\bibnamefont {König}}, \bibinfo {author} {\bibfnamefont
  {C.}~\bibnamefont {Ames}}, \bibinfo {author} {\bibfnamefont {H.}~\bibnamefont
  {Buhmann}}, \bibinfo {author} {\bibfnamefont {P.}~\bibnamefont {Leubner}},
  \bibinfo {author} {\bibfnamefont {L.~W.}\ \bibnamefont {Molenkamp}}, \bibinfo
  {author} {\bibfnamefont {S.-C.}\ \bibnamefont {Zhang}}, \bibinfo {author}
  {\bibfnamefont {D.}~\bibnamefont {Goldhaber-Gordon}}, \bibinfo {author}
  {\bibfnamefont {M.~A.}\ \bibnamefont {Kelly}},\ and\ \bibinfo {author}
  {\bibfnamefont {Z.-X.}\ \bibnamefont {Shen}},\ }\bibfield  {title} {\bibinfo
  {title} {Unexpected edge conduction in mercury telluride quantum wells under
  broken time-reversal symmetry},\ }\href {https://doi.org/10.1038/ncomms8252}
  {\bibfield  {journal} {\bibinfo  {journal} {Nature Communications}\ }\textbf
  {\bibinfo {volume} {6}},\ \bibinfo {pages} {7252} (\bibinfo {year}
  {2015})}\BibitemShut {NoStop}%
\bibitem [{\citenamefont {Wan}\ \emph {et~al.}(2011)\citenamefont {Wan},
  \citenamefont {Turner}, \citenamefont {Vishwanath},\ and\ \citenamefont
  {Savrasov}}]{wan2011}%
  \BibitemOpen
  \bibfield  {author} {\bibinfo {author} {\bibfnamefont {X.}~\bibnamefont
  {Wan}}, \bibinfo {author} {\bibfnamefont {A.~M.}\ \bibnamefont {Turner}},
  \bibinfo {author} {\bibfnamefont {A.}~\bibnamefont {Vishwanath}},\ and\
  \bibinfo {author} {\bibfnamefont {S.~Y.}\ \bibnamefont {Savrasov}},\
  }\bibfield  {title} {\bibinfo {title} {Topological semimetal and fermi-arc
  surface states in the electronic structure of pyrochlore iridates},\ }\href
  {https://doi.org/10.1103/PhysRevB.83.205101} {\bibfield  {journal} {\bibinfo
  {journal} {Phys. Rev. B}\ }\textbf {\bibinfo {volume} {83}},\ \bibinfo
  {pages} {205101} (\bibinfo {year} {2011})}\BibitemShut {NoStop}%
\bibitem [{Sup()}]{SuppMat}%
  \BibitemOpen
  \href@noop {} {}\bibinfo {note} {See Supplementary materials for details on
  realisation of robust edge conduction from WN\textsubscript{F}s, and results
  for Zeeman field terms dependent on orbital field strength.}\BibitemShut
  {Stop}%
\bibitem [{\citenamefont {Dantscher}\ \emph {et~al.}(2017)\citenamefont
  {Dantscher}, \citenamefont {Kozlov}, \citenamefont {Scherr}, \citenamefont
  {Gebert}, \citenamefont {B\"arenf\"anger}, \citenamefont {Durnev},
  \citenamefont {Tarasenko}, \citenamefont {Bel'kov}, \citenamefont
  {Mikhailov}, \citenamefont {Dvoretsky}, \citenamefont {Kvon}, \citenamefont
  {Ziegler}, \citenamefont {Weiss},\ and\ \citenamefont
  {Ganichev}}]{dantscher2017}%
  \BibitemOpen
  \bibfield  {author} {\bibinfo {author} {\bibfnamefont {K.-M.}\ \bibnamefont
  {Dantscher}}, \bibinfo {author} {\bibfnamefont {D.~A.}\ \bibnamefont
  {Kozlov}}, \bibinfo {author} {\bibfnamefont {M.~T.}\ \bibnamefont {Scherr}},
  \bibinfo {author} {\bibfnamefont {S.}~\bibnamefont {Gebert}}, \bibinfo
  {author} {\bibfnamefont {J.}~\bibnamefont {B\"arenf\"anger}}, \bibinfo
  {author} {\bibfnamefont {M.~V.}\ \bibnamefont {Durnev}}, \bibinfo {author}
  {\bibfnamefont {S.~A.}\ \bibnamefont {Tarasenko}}, \bibinfo {author}
  {\bibfnamefont {V.~V.}\ \bibnamefont {Bel'kov}}, \bibinfo {author}
  {\bibfnamefont {N.~N.}\ \bibnamefont {Mikhailov}}, \bibinfo {author}
  {\bibfnamefont {S.~A.}\ \bibnamefont {Dvoretsky}}, \bibinfo {author}
  {\bibfnamefont {Z.~D.}\ \bibnamefont {Kvon}}, \bibinfo {author}
  {\bibfnamefont {J.}~\bibnamefont {Ziegler}}, \bibinfo {author} {\bibfnamefont
  {D.}~\bibnamefont {Weiss}},\ and\ \bibinfo {author} {\bibfnamefont {S.~D.}\
  \bibnamefont {Ganichev}},\ }\bibfield  {title} {\bibinfo {title}
  {Photogalvanic probing of helical edge channels in two-dimensional hgte
  topological insulators},\ }\href {https://doi.org/10.1103/PhysRevB.95.201103}
  {\bibfield  {journal} {\bibinfo  {journal} {Phys. Rev. B}\ }\textbf {\bibinfo
  {volume} {95}},\ \bibinfo {pages} {201103} (\bibinfo {year}
  {2017})}\BibitemShut {NoStop}%
\bibitem [{\citenamefont {de~Juan}\ \emph {et~al.}(2017)\citenamefont
  {de~Juan}, \citenamefont {Grushin}, \citenamefont {Morimoto},\ and\
  \citenamefont {Moore}}]{de_juan_quantized_2017}%
  \BibitemOpen
  \bibfield  {author} {\bibinfo {author} {\bibfnamefont {F.}~\bibnamefont
  {de~Juan}}, \bibinfo {author} {\bibfnamefont {A.~G.}\ \bibnamefont
  {Grushin}}, \bibinfo {author} {\bibfnamefont {T.}~\bibnamefont {Morimoto}},\
  and\ \bibinfo {author} {\bibfnamefont {J.~E.}\ \bibnamefont {Moore}},\
  }\bibfield  {title} {\bibinfo {title} {Quantized circular photogalvanic
  effect in {Weyl} semimetals},\ }\href {https://doi.org/10.1038/ncomms15995}
  {\bibfield  {journal} {\bibinfo  {journal} {Nature Communications}\ }\textbf
  {\bibinfo {volume} {8}},\ \bibinfo {pages} {15995} (\bibinfo {year}
  {2017})}\BibitemShut {NoStop}%
\end{thebibliography}%

\pagebreak

\appendix

\clearpage

\makeatletter
\renewcommand{\theequation}{S\arabic{equation}}
\renewcommand{\thefigure}{S\arabic{figure}}
\renewcommand{\thesection}{S\arabic{section}}
\setcounter{equation}{0}
\setcounter{section}{0}
\onecolumngrid


\begin{center}
  \textbf{\large Supplemental material for ``Signatures of the quantum skyrmion Hall effect in the Bernevig-Hughes-Zhang model''}\\[.2cm]
 Reyhan Ay$^{1,2}$, Adipta Pal$^{1,2}$ and Ashley M. Cook$^{1,2,*}$\\[.1cm]
  {\itshape ${}^1$Max Planck Institute for Chemical Physics of Solids, Nöthnitzer Strasse 40, 01187 Dresden, Germany\\
  ${}^2$Max Planck Institute for the Physics of Complex Systems, Nöthnitzer Strasse 38, 01187 Dresden, Germany\\}
  ${}^*$Electronic address: cooka@pks.mpg.de\\
(Dated: \today)\\[1cm]
\end{center}

\section{State preparation for robust edge conduction}

\subsection{State preparation in the presence of all Zeeman field components}
The Bernevig-Hughes-Zhang (BHZ) model with inversion symmetry breaking atomic spin orbit coupling (SOC) term $\mathcal{H}_{IB}$ and Zeeman field term $\mathcal{H}_Z$ coupling to the spin degree of freedom (DOF), $\tau$, is
\begin{align}
&\mathcal{H}_{BHZ}(k_x,k_y^\prime) = \mathcal{H}_0(k_x,k_y^\prime)+{H}_{IB}+\mathcal{H}_Z,
\end{align}
where
\begin{align}
&\mathcal{H}_0(k_x,k_y^\prime)=(u+\cos(k_x)+\cos(k_y^\prime))\tau^0\sigma^z+\sin(k_x)\tau^z\sigma^x+\sin(k_y^\prime)\tau^0\sigma^y\\ 
&\mathcal{H}_{IB}=c_A\tau^x\sigma^y\\ 
&\mathcal{H}_Z= \boldsymbol{h}\cdot\boldsymbol{\tau}\sigma^0= g\mathbf{B}\cdot\boldsymbol{\tau}\sigma^0.
\end{align}
Here, the gauge fixing condition for the Peierls substitution is given as, $k_y\rightarrow k_y^\prime = k_y+eB_zx$. Note that only the out of plane magnetic field can contribute to the orbital magnetization.\\

\begin{center}
\begin{figure}[htb!]
    \centering
    \includegraphics[width=0.7\textwidth]{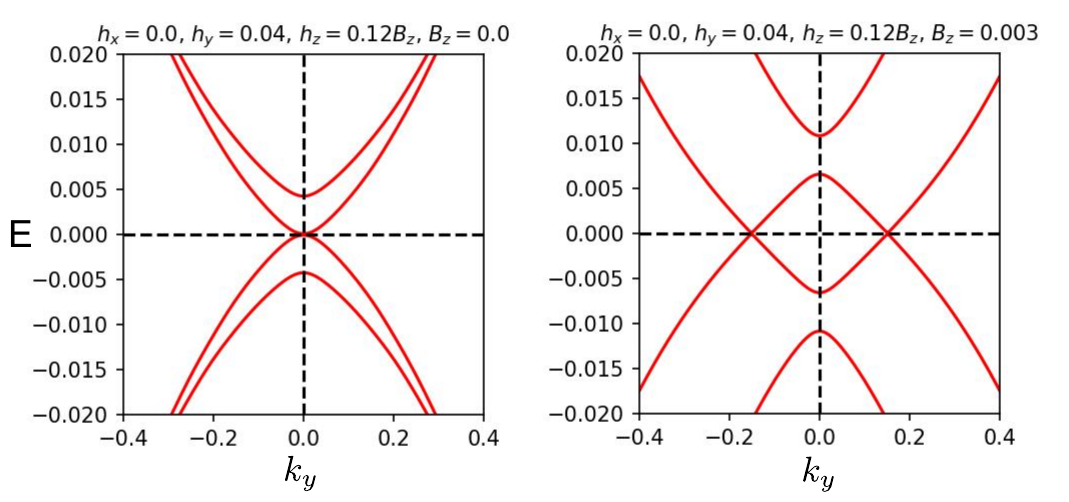}
    \caption{Slab spectra vs. $k_y$ for OBC in x($L_x=101$) for parameter values, $u=-1.56$, $c_A=-0.77$, $h_x=0$, $h_y=0.04$, and $h_z=0.12B_z$ with (a) $B_z=0.0$ and (b) $B_z=0.003$. }
    \label{fig:bhztilt0}
\end{figure}
\end{center}

We will consider cases in which $\boldsymbol{h}=g\mathbf{B}=(h_x,h_y,h_z)$ with each of $h_x$, $h_y$, and $h_z$ finite. Starting with a time-reversal symmetric system, we first show $h_y$ removes the Kramers degeneracy. If we then introduce finite $h_z$ and orbital magnetization $B_z$, bands close to zero energy can then furthermore intersect to form gapless points in the edge spectrum at zero energy. We illustrate this in Fig.~\ref{fig:bhztilt0}(a) and (b) before and after applying $B_z$  respectively, for the parameter set used for Fig.~4 in the main text. We show a more general prescription for realising such features in the slab spectra later in the supplementary materials.\\

Starting from such a value of $h_y$, with $h_z$ and $B_z$ each zero, and then introducing finite $h_x$ instead shifts the quadratic bands in opposite directions along $k_y$ while preserving a band-touching at $k_y=0$, such that a Fermi level at zero-energy intersects two bands at finite $k_y$ as well as $k_y=0$. Introducing finite $h_z$ and $B_z$ then generically yields a finite minimum direct gap, but also an overall negative indirect gap in the edge spectrum, such that the Fermi level is guaranteed to intersect edge bands regardless of placement within the bulk energy gap.

We can identify the pair of gapless zero-energy points in the slab spectrum for finite $h_y$, $h_z$, and $B_z$ but zero $h_x$, with the pair of small finite direct gaps realised at finite $k_y$ for $h_x$ instead also finite. Thus, interplay between finite $h_x$, $h_y$, $h_z$, and $B_z$ yields an overall negative minimum indirect gap, while also introducing a much smaller minimum direct energy gap in the slab spectrum corresponding to gapping out of  severely-fuzzified Weyl nodes WN\textsubscript{F}s.

\begin{center}
\begin{figure}[htb!]
    \centering
    \includegraphics[width=0.95\textwidth]{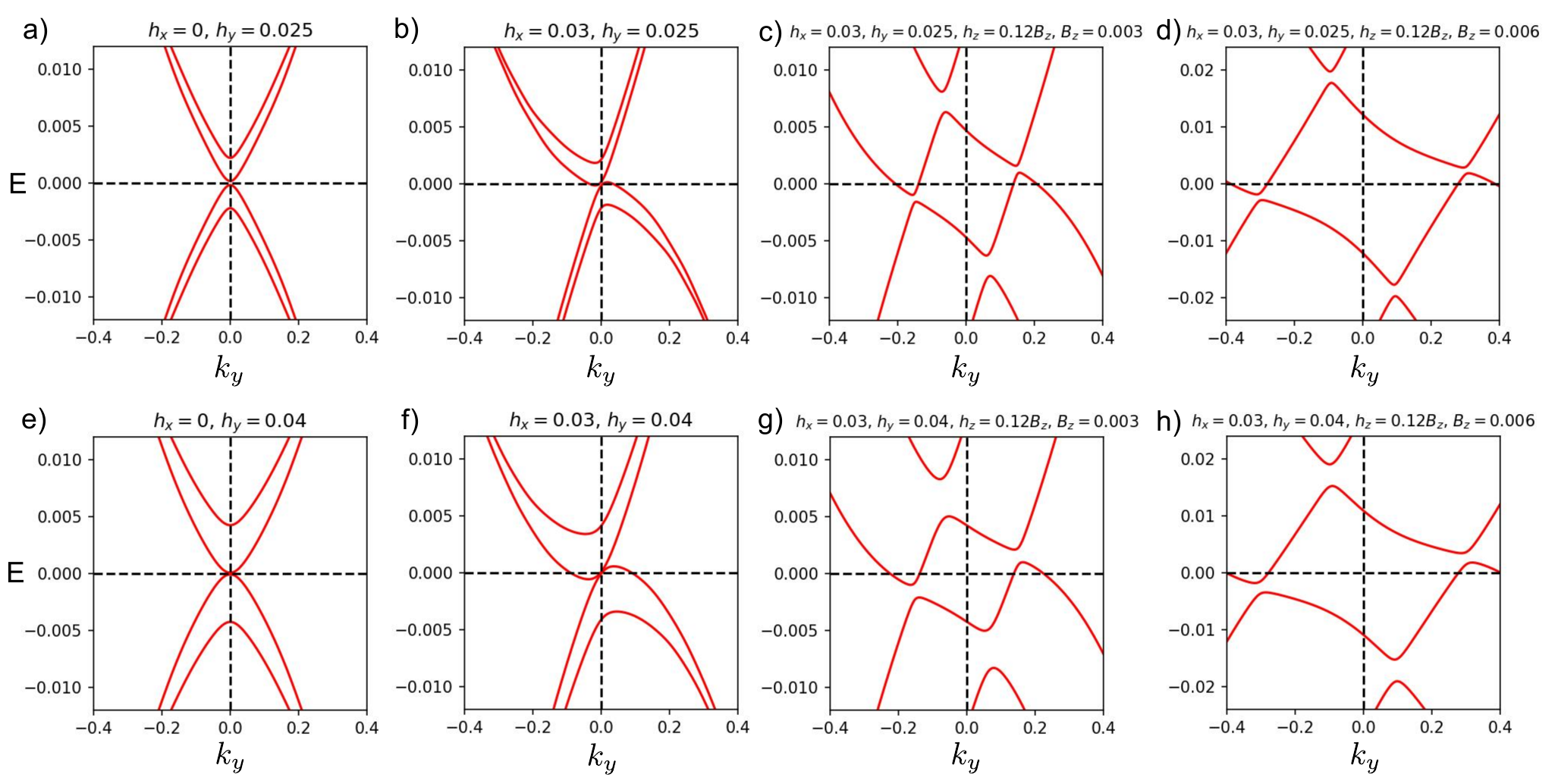}
    \caption{Slab spectra vs. $k_y$ for OBC in x ($L_x=101$) in the BHZ model with parameters $u=-1.56$, $c_A=-0.77$ with (a) $\boldsymbol{h}=(0.0,0.025,0.0)$, (b) $\boldsymbol{h}=(0.03,0.025,0.0)$, (c) $\boldsymbol{h}=(0.03,0.025,0.12B_z)$, $B_z=0.003$, (d) $\boldsymbol{h}=(0.03,0.025,0.12B_z)$, $B_z=0.006$, (e) $\boldsymbol{h}=(0.0,0.04,0.0)$, (f) $\boldsymbol{h}=(0.03,0.04,0.0)$, (g) $\boldsymbol{h}=(0.03,0.04,0.12B_z)$, $B_z=0.003$, (h) $\boldsymbol{h}=(0.03,0.04,0.12B_z)$, $B_z=0.006$. Subfigure (g) is exactly the case outlined in Fig. 4 in the main text.}
    \label{fig:bhztilt1}
\end{figure}
\end{center}

We illustrate this phenomena step by step in Fig.~\ref{fig:bhztilt1} for the parameter set related to Fig. 4 in the main text. At that given value of $c_A$, introducing Zeeman field along y, $h_y$ gaps out the initial degeneracy of bands as shown in Fig.~\ref{fig:bhztilt1}(a) and (e) for $h_y=0.025$ and $h_y=0.04$ respectively, where comparison shows the hybridization gap between the initially degenerate bands is larger for larger value of $h_y$. Introducing $h_x=0.03$ to both (a) and (e) leads to Fig.~\ref{fig:bhztilt1}(b) and (f) where the quadratic bands for positive and negative energy are shifted from $k_y=0$ in opposite directions and intersect the horizontal axis. In Fig.~\ref{fig:bhztilt1}(c) and (g), we further introduce Zeeman coupling along z, $h_z=0.12B_z$ and orbital magnetization $B_z=0.003$ to (b) and (f) which leads to tilted and gapped out WN\textsubscript{F}s. The position of the gapped out nodes are determined by $h_z$ and $B_z$ as can be seen by comparing Fig.~\ref{fig:bhztilt1}(d) with (c) and Fig.~\ref{fig:bhztilt1}(g) with (h).

\begin{center}
\begin{figure}[htb!]
    \centering
    \includegraphics[width=0.65\textwidth]{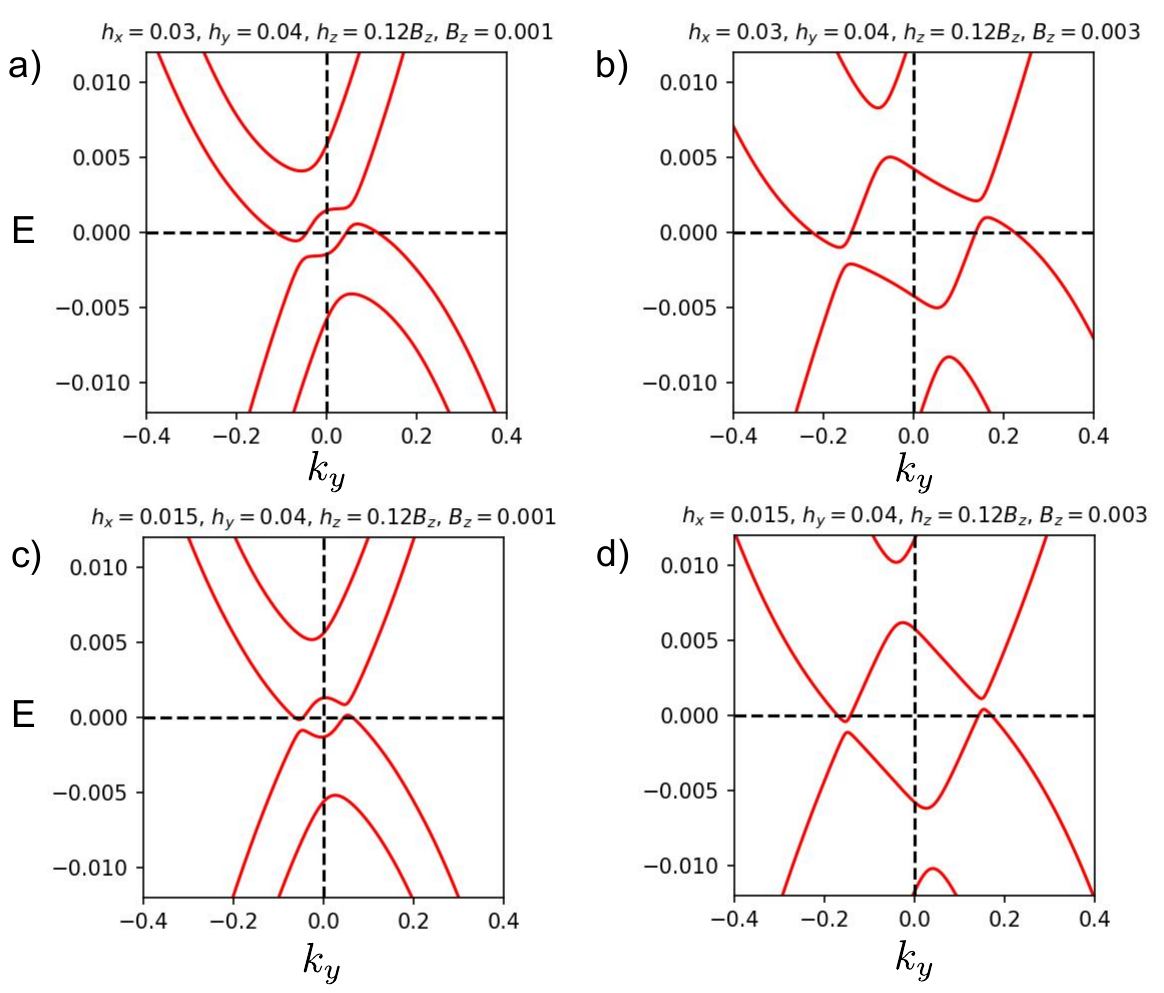}
    \caption{Slab spectra vs. $k_y$ for OBC in x $(L_x=101)$, with parameters $u=-1.56$, $c_A=-0.77$, $h_y=0.04$ for changing $h_x$ vertically and changing $B_z$ horizontally. (a) $h_x=0.03$, $h_z=0.12B_z$, $B_z=0.001$, (b) $h_x=0.03$, $h_z=0.12B_z$, $B_z=0.003$, (c) $h_x=0.015$, $h_z=0.12B_z$, $B_z=0.001$ and (d) $h_x=0.015$, $h_z=0.12B_z$, $B_z=0.003$.}
    \label{fig:bhztilt2}
\end{figure}
\end{center}

The direct gaps opened at locations of severely-fuzzified Weyl nodes WN\textsubscript{F} gapless points increases with the magnitude of $h_x$ as can be shown by comparing Fig.~\ref{fig:bhztilt2}(a) and (c) for $h_x=0.03$ and $h_x=0.015$ respectively. We also show the indirect band gap in the OBC slab spectra depends on $B_z$ as mentioned previously in Fig.~\ref{fig:bhztilt2}(b) with $B_z=0.001$ compared with (a) with $B_z=0.003$ and Fig.~\ref{fig:bhztilt2}(d) compared with (c) with equivalent change in $B_z$.

\begin{center}
\begin{figure}[htb!]
    \centering
    \includegraphics[width=0.75\textwidth]{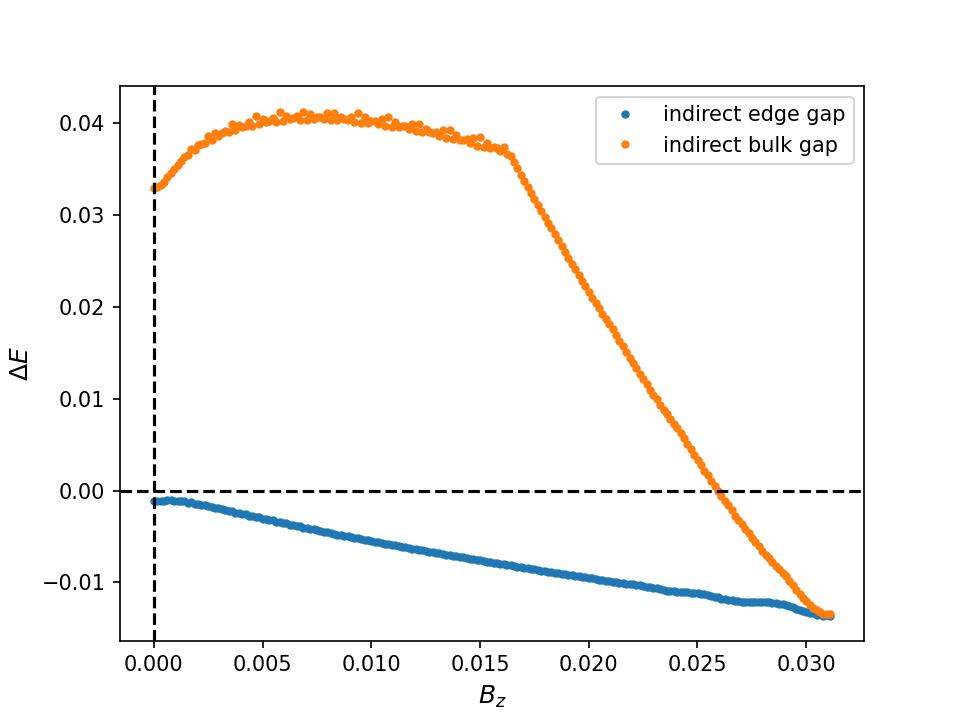}
    \caption{Minimum indirect gap between valence and conduction bands for states in the bulk or localised on the boundary, in the slab spectra for OBCs in the $\hat{x}$-direction and PBCs in the $\hat{y}$-direction. System size is taken to be $L_x=101$ along x for parameters $u=-1.56$, $c_A=-0.77$, $h_x=0.03$, $h_y=0.04$, $h_z=0.12B_z$. The minimum indirect band gap for edge bands is negative, becoming more negative as $B_z$ is increased. The minimum indirect gap for bulk states is positive up to a large value of $B_z$ relative to $B_c$ in the main text.}
    \label{fig:bhzindirectgap}
\end{figure}
\end{center}

This minimum negative indirect band gap in the edge spectrum persists and increases when $B_z$ is increased as shown in Fig.~\ref{fig:bhzindirectgap}. 

\section{Slab spectra for open boundary conditions in the y direction}
\begin{center}
\begin{figure}[htb!]
    \centering
    \includegraphics[width=0.7\textwidth]{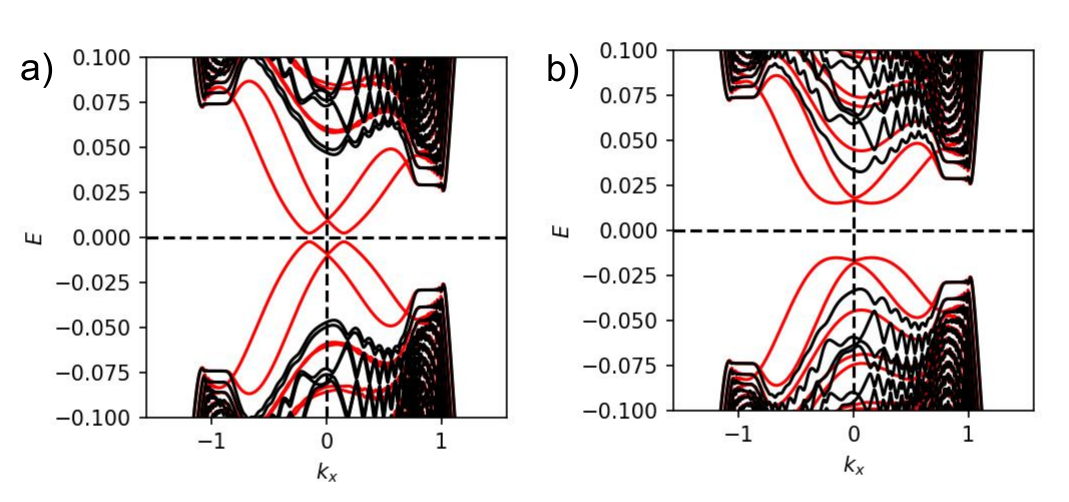}
    \caption{Slab spectra vs. $k_x$ for OBC (red) and PBC(black) in y($L_y=101$) with parameters $u=-1.56$, $c_A=-0.77$, $h_z=0.12B_z$, $B_z=0.003$ for (a) $h_x=0.0$, $h_y=0.04$ and (b) $h_x=0.015$, $h_y=0.04$.}
    \label{fig:bhztilt3}
\end{figure}
\end{center}

We show the slab spectra for OBCs in the $\hat{y}$-direction and PBCs in the $\hat{x}$-direction, as a function of $k_x$ in Fig~\ref{fig:bhztilt3}: for similar parameter ranges as in Fig.~2 and Fig.~4 in the main text, we show, in Fig.~\ref{fig:bhztilt3}(a) and (b), that there exists a positive minimum indirect band gap in the edge spectra, i.e., the Fermi level is not guaranteed to intersect edge bands, meaning local density of states is not generally expected to be significant on these edges. Fig.~\ref{fig:bhztilt3}(a) and (b) show slab spectra for these boundary conditions for two parameter sets differing by their respective values of $h_x$, illustrating how $h_x$ enlarges the positive minimum indirect gap in the edge spectrum.

\newpage
\section{State preparation for compactified Weyl nodes in the slab spectra of the BHZ model}

Here, we present some methods of realising severely-fuzzified Weyl nodes WN\textsubscript{F}s more generically, starting from helical edge states of the BHZ model. We focus on cases in which the Zeeman field term is proportional to orbital magnetic field strength as considered in Ma~\emph{et al.}~\cite{ma_unexpected_2015}. 
\begin{center}
\begin{figure}[h!]
    \centering
    \includegraphics[width=.95\textwidth]{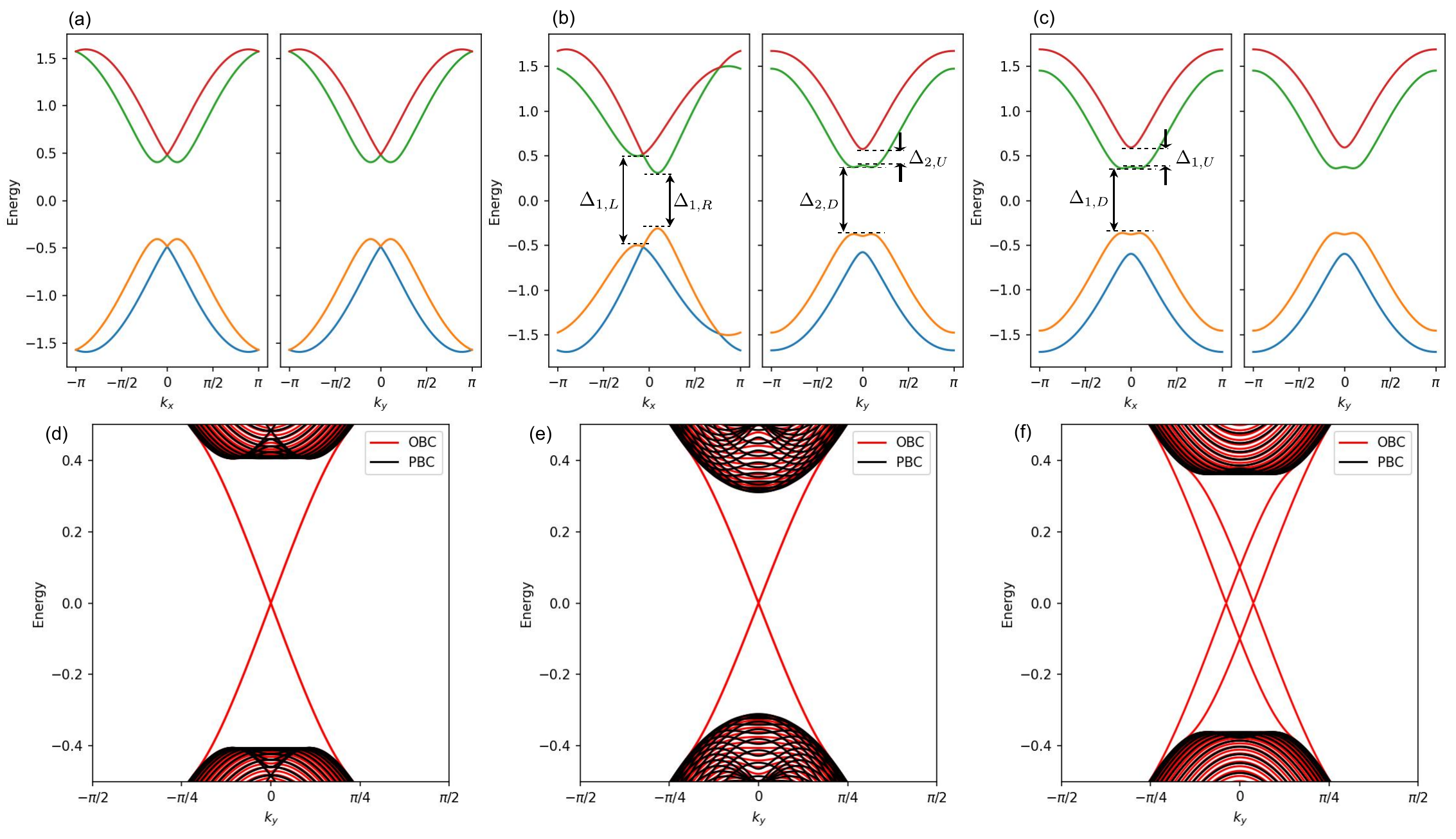}
    \caption{Bulk dispersion vs. $k_x$ at $k_y=0$ and vs. $k_y$ at $k_x=0$ for (a)$u=-1.56$, $c_A=-0.2$, $h_x=0=h_z$, (b) $u=-1.56$, $c_A=-0.2$, $h_y=0.1$, $gB_z=0$, and (c) $u=-1.56$, $c_A=-0.2$, $h_y=0$, $h_z=0.12$ (Zeeman field is oriented in the z direction, orbital magnetic field is trivial). In (d), (e) and (f) we plot the corresponding slab spectra for (a), (b) and (c) respectively.}
    \label{fig:bhzbulkslab1}
\end{figure}
\end{center}
We first consider the bulk dispersion in the absence of orbital magnetization but with finite Zeeman field, where we will investigate six different energy gaps---$\Delta_{1,L}$, $\Delta_{1,R}$, $\Delta_{2,D}$, $\Delta_{2,U}$, $\Delta_{1,D}$, $\Delta_{1,U}$---as shown in Fig.~\ref{fig:bhzbulkslab1}. In Fig.~\ref{fig:bhzbulkslab1}(a), we show the bulk dispersion of the BHZ model vs. $k_x$ at $k_y=0$ (left) and vs. $k_y$ at $k_x=0$ (right) for non-negligible atomic SOC. In Fig.~\ref{fig:bhzbulkslab1}(b), we show the effect of additionally applying a Zeeman field in the y direction. There are two different band gaps, $\Delta_{1,L}$ and $\Delta_{1,R}$, in the bulk spectrum vs. $k_x$. There are also two gaps, $\Delta_{2,U}$ and $\Delta_{2,D}$, in the bulk spectrum vs. $k_y$. In Fig.~\ref{fig:bhzbulkslab1}(c), we introduce a Zeeman field in the z direction, which yields band gaps $\Delta_{1,U}$ and $\Delta_{1,D}$ between bands indexed 3 and 4, and 2 and 3, in energy, respectively, vs. $k_x$. In this case, $\Delta_{1,U}$ and $\Delta_{1,D}$ are equal to $\Delta_{2,U}$ and $\Delta_{2,D}$, respectively, for the bulk dispersion vs. $k_y$. Zeeman field oriented in the z direction corresponds to all gaps of interest being finite with the relations, $\Delta_{1,D}=\Delta_{1,L}=\Delta_{1,R}=\Delta_{2,D}$ and $\Delta_{1,U}=\Delta_{2,U}$. 

\begin{center}
\begin{figure}[h!]
    \centering
    \includegraphics[width=0.95\textwidth]{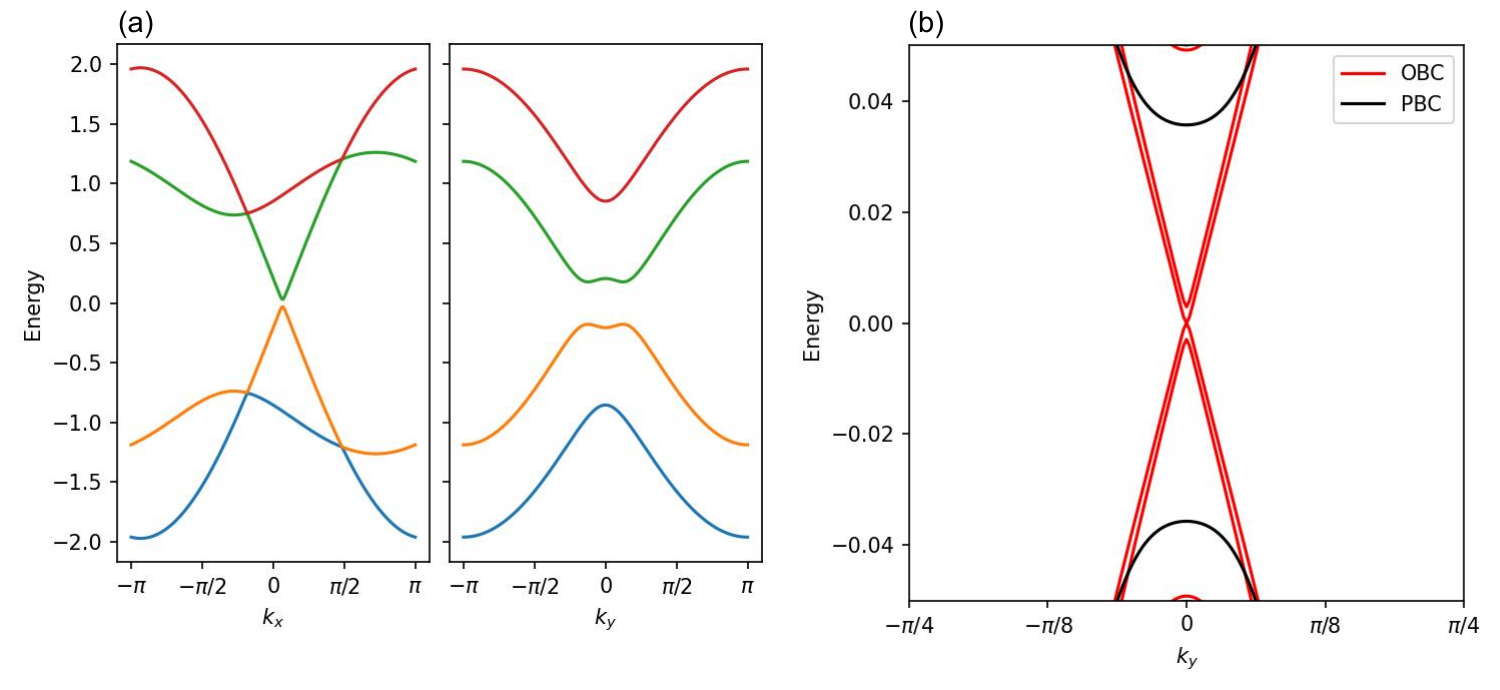}
    \caption{(a)Bulk dispersion vs. $k_x$ at $k_y=0$ and vs. $k_y$ at $k_x=0$ for $u=-1.56$, $c_A=-0.2$, $h_y=0.39$, $h_z=0$. (b) Corresponding slab spectra vs. $k_y$ for open(red) and periodic(black) boundary in the x direction, with system size $L_x=101$. We observe a higher order band touching at $k_y=0$.}
    \label{fig:bhzbulkslab2}
\end{figure}
\end{center}

\begin{center}
\begin{figure}[h!]
    \centering
    \includegraphics[width=0.7\textwidth]{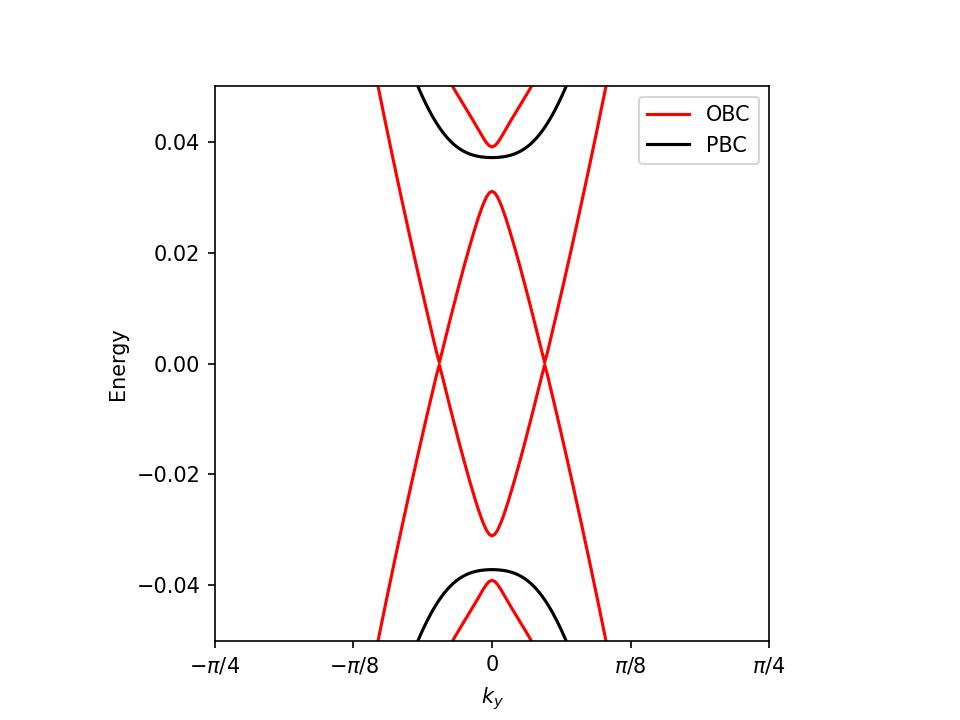}
    \caption{Slab spectra vs. $k_y$ for OBCs in the $\hat{x}$-direction, with system size $L_x=101$ for the parameter values, $u=-1.56$, $c_A=-0.2$, $h_y=0.39$, $h_z=0.12B_z$, $B_z=0.003$.}
    \label{fig:bhzbulkslab3}
\end{figure}
\end{center}

A notable feature of the slab spectrum relevant to state preparation is removal of band degeneracy. First, consider the slab spectrum shown in Fig.~\ref{fig:bhzbulkslab1}(d) in the absence of an applied Zeeman field component in the z direction, and also in the slab spectrum in Fig.~\ref{fig:bhzbulkslab1}(e) for finite Zeeman field oriented in the y direction, below the threshold value of $B_c$, when we are in a regime of negligible hybridisation gap $\Delta$ according to Fig.~2 of the main text, as shown in Fig.~\ref{fig:bhzbulkslab1}(f). Beyond a critical value of Zeeman field strength for Zeeman field oriented in the y direction, when $\Delta_{1,R}<\Delta_{1,L}$ as shown in the bulk dispersion in Fig.~\ref{fig:bhzbulkslab2}(a), we instead observe loss of  band degeneracy in the slab spectrum vs. $k_y$ within the bulk gap, for OBCs in the $\hat{x}$-direction, shown in Fig.~\ref{fig:bhzbulkslab2}(b). From this scenario, we can introduce a magnetic field in the z direction, which contributes to the orbital magnetization by coupling to the momenta as well as to the Zeeman field component in the z direction. We show the corresponding slab spectra vs. $k_y$ in Fig.~\ref{fig:bhzbulkslab3}, which exhibits two nodes at finite $k_y$ which we can compare with Fig.~\ref{fig:bhzbulkslab1}(f) where only the Zeeman field component in the $\hat{z}$-direction is finite. The Zeeman field component along y, $h_y$, is responsible for producing the gap which in turn isolates the pair of gapless points at zero energy from other bands. Therefore, if the applied magnetic field has a strong enough in-plane component in the y direction to generally eliminate band degeneracy, a finite out-of-plane field component then yields edge states with gapless points in the slab spectrum. Such points yield signatures in local density of states corresponding to edge conduction even for strong breaking of time-reversal symmetry, similarly to the findings of Ma~\emph{et al.}~\cite{ma_unexpected_2015} and results shown in Fig.~4 in the main text. 

\end{document}